\documentclass[aps,prb,reprint,superscriptaddress,showpacs]{revtex4-1}
\usepackage{graphicx}
\usepackage{natbib}
\usepackage{amsmath}
\usepackage{amssymb}
\usepackage{amsthm}
\usepackage{bm}
\usepackage{amsmath}
\usepackage{amssymb}
\usepackage{amsthm}
\usepackage{graphicx}
\usepackage{empheq}
\usepackage{color}
\usepackage{float}

\newcommand{\bra}[1]{\left<#1\right|} 
\newcommand{\ket}[1]{\left|#1\right>} 
\newcommand{\braket}[2]{\left<#1|#2\right>}

\begin{document}

%\preprint{}

\title{First-principles hyperfine tensors for electrons and holes in GaAs and silicon}


\author{Pericles Philippopoulos}
%\thanks{}
\affiliation{Department of Physics, McGill University, 3600 rue University, Montreal, Qc H3A 2T8, Canada}

\author{Stefano Chesi}
\affiliation{Beijing Computational Science Research Center, Beijing 100193, China}
\affiliation{Department of Physics, Beijing Normal University, Beijing 100875, China}

\author{W. A. Coish}
\affiliation{Department of Physics, McGill University, 3600 rue University, Montreal, Qc H3A 2T8, Canada}


\date{\today}

\begin{abstract}
Understanding (and controlling) hyperfine interactions in semiconductor nanostructures is important for fundamental studies of material properties as well as for quantum information processing with electron, hole, and nuclear-spin states. 
Through a combination of first-principles density-functional theory (DFT) and $\mathbf{k}\cdot\mathbf{p}$ theory, we have calculated hyperfine tensors for electrons and holes in GaAs and crystalline silicon.  
Accounting for relativistic effects near the nuclear core, we find contact hyperfine interactions for electrons in GaAs that are consistent with Knight-shift measurements performed on GaAs quantum wells and are roughly consistent with prior estimates extrapolated from measurements on InSb.  We find that a combination of DFT and $\mathbf{k}\cdot\mathbf{p}$ theory (DFT+$\mathbf{k}\cdot\mathbf{p}$) is necessary to accurately determine the contact hyperfine interaction for electrons at a conduction-band minimum in silicon that is consistent with bulk Knight-shift measurements.
For hole spins in GaAs, the overall magnitude of the hyperfine couplings we find from DFT is consistent with previous theory based on free-atom properties, and with heavy-hole Overhauser shifts measured in GaAs (and InGaAs) quantum dots.
In addition, we theoretically predict that the heavy-hole hyperfine coupling to the As nuclear spins is stronger and almost purely Ising, while the (weaker) coupling to the Ga nuclear spins has significant non-Ising corrections. 
In the case of hole spins in silicon, we find (surprisingly) that the strength of the hyperfine interaction in the valence band is comparable to that in the conduction band and that the hyperfine tensors are highly anisotropic (Ising) in the heavy-hole subspace. These results suggest that the hyperfine coupling cannot be ruled out as a limiting mechanism for coherence ($T_2^{\ast}$) times recently measured for heavy holes in silicon quantum dots.

\end{abstract}
%\pacs{31.30.Gs, 73.22.-f, 75.75.-c}
%31.30.Gs (hyperfine interactions); 
%73.22.-f (electronic structure->condensed matter->nanoscale materials); 
%75.75.-c (quantum dots->magnetic properties of)

\maketitle

\section{Introduction}

Semiconductor nanostructures are essential to confine spin qubits in quantum dots\cite{kloeffel2013prospects,eriksson2004spin} and to implement other spintronic devices.\cite{awschalom2013quantum}  From the perspective of quantum transport and low electronic noise, a near-ideal platform for these devices is provided by high-mobility heterostructures based on GaAs.\cite{kane1995high,seamons2007undoped}  However, every stable isotope of Ga and As has a finite nuclear spin, resulting in a coupling of the electron (or hole) spins to a large reservoir of nuclear spins through the hyperfine interaction.\cite{merkulov2002electron,khaetskii2002electron,warburton2013single}  If the hyperfine interaction is not fully understood and controlled, this interaction may lead to a randomization of the spins in spintronic or spin-qubit devices.  To avoid the effects of the strong hyperfine interactions for electrons in GaAs, there have been many recent studies of alternative devices based on electron spins in silicon, for which the majority isotope has no nuclear spin or based on hole spins in either GaAs or silicon, for which the hyperfine couplings are weak.   

A key advantage of hole spins over electron spins in GaAs is that holes have a weaker hyperfine coupling.\cite{fischer2008spin, fallahi2010measurement, chekhovich2013element, vidal2016hyperfine, prechtel2016decoupling} 
Because the hole hyperfine interaction is anisotropic, it may be possible to further reduce or eliminate the effects of the hole hyperfine coupling through motional-averaging.\cite{fischer2008spin,wang2012spin, chesi2014controlling, wang2015maximizing} 
An additional benefit of hole spins over electrons is a stronger spin-orbit coupling, leading to robust all-electric hole-spin manipulation.\cite{bulaev2007electric,pribiag2013electrical,maurand2016cmos, bogan2018landau,studenikin2019electrically}
This advantage afforded by a stronger spin-orbit coupling does not necessarily come at the cost of significantly shorter spin-relaxation ($T_1$) times in confined nanostructures.\cite{bulaev2005spin,heiss2007observation}
Despite these advantages, the electrical instability of p-doped GaAs nanostructures\cite{zailer1994phase, grbic2005single, burke2012origin} has made experimental investigations of these systems difficult. 
Recent advances in fabricating few-hole quantum dots from undoped samples\cite{tracy2014few} have now opened up a greater range of possibilities for hole-spin devices. 
Undoped devices have shown Pauli spin blockade,\cite{wang2016anisotropic, bogan2017consequences, bogan2019single} and measurements have been performed revealing hole-spin relaxation times ($T_1$),\cite{bogan2019single} $g$-factors,\cite{bogan2017consequences} and spin-orbit couplings.\cite{marcellina2018electrical}  Despite these advances, many details of the hyperfine couplings for holes in GaAs and silicon remain largely unknown.

Electron spins in silicon quantum dots have now reached a level of control and coherence that makes them serious contenders for elements of near-future quantum processors.\cite{watson2018programmable,zajac2018resonantly,fogarty2018integrated,huang2019fidelity} 
Because of the small abundance ($\sim 4.7\%$) of spinful $^{29}\mathrm{Si}$ nuclei in natural silicon,  electron (and hole) spins in silicon nanostructures interact more weakly with the nuclear-spin bath. Coherence times for electron spins in natural silicon quantum dots are nevertheless often limited by the hyperfine interaction.\cite{maune2012coherent} 
Isotopically purified $^{28}\mathrm{Si}$ has been used as an alternative nuclear-spin free host,\cite{saeedi2013room, muhonen2014storing, veldhorst2014addressable, veldhorst2015two, eng2015isotopically, sigillito2019coherent}
but even in these systems, the hyperfine coupling to the few remaining (residual) $^{29}\mathrm{Si}$ nuclear spins can have a measurable effect on a quantum-dot-bound electron-spin.\cite{zhao2019single, hensen2019silicon}

Removing the nuclear spins from the host material suppresses decoherence, but it also precludes the potential benefits of a finite hyperfine interaction. 
These benefits include addressing the nuclear spins and using them as additional qubits for a quantum register\cite{wolfowicz201629si, bradley2019ten} or a quantum memory,\cite{taylor2003long, pla2014coherent, bradley2019ten} and using the nuclear spins to apply local effective magnetic fields on the electron or hole spins to locally manipulate them.\cite{laird2007hyperfine, foletti2009universal}
It is therefore important to understand the strength and properties of the hyperfine Hamiltonian for electron and hole spins in semiconductor nanostructures.
This knowledge could allow negative effects to be suppressed while maintaining potential advantages of the coupling with the nuclear spins. 

Knowing material-specific hyperfine parameters is also important or required to interpret measurements of physical quantities. 
These quantities include the degree of nuclear polarization from Overhauser shift measurements,\cite{dobers1988electrical, baugh2007large, latta2009confluence, gangloff2019quantum} and the nuclear spin polarization in the quantum Hall regime.\cite{tycko1995electronic, kuzma1998ultraslow, khandelwal1998optically, smet2002gate, tiemann2012unraveling}

The goal of this work is to accurately calculate the hyperfine parameters for electrons and holes in GaAs and silicon.
Earlier attempts at calculating hyperfine constants have relied on estimates of the electronic density (or wave function) based on non-relativistic free-atom properties such as the free-atom orbital radius.\cite{paget1977low, fischer2008spin,testelin2009hole, machnikowski2019hyperfine} 
Instead, here we calculate the hyperfine parameters using all-electron density-functional theory (DFT) accompanied by $\mathbf{k}\cdot\mathbf{p}$ theory (DFT+$\mathbf{k}\cdot\mathbf{p}$), accounting for relativistic effects, and fully including the anisotropic crystalline environment in our analysis.
Typically, DFT procedures are used to calculate electronic densities. 
If the electronic states under consideration can be approximated as uncorrelated product states of spin and orbital degrees of freedom, the density alone is sufficient to calculate the hyperfine parameters.\cite{vandewalle1993first} 
This approach has been used to calculate hyperfine parameters for electrons in silicon.\cite{assali2011hyperfine}
However, this procedure cannot generally be applied to states (such as the valence-band states of GaAs and silicon) where the spin-orbit coupling is relevant and the states are therefore not necessarily product states. 
Moreover, the density alone provides no information about the phase of the wave function.
Thus, e.g., matrix elements of the angular momentum operator cannot generally be calculated from the density alone and the nuclear-orbital interaction [$\sim \mathbf{L}\cdot \bm{I}$, see Eq.~\eqref{eq:hrSingleParticleMatrix} below] is often neglected.\cite{feller1984ab, vandewalle1993first, yazyev2008hyperfine, gali2008ab, ghosh2019all}
In contrast, here we apply DFT to evaluate the Kohn-Sham orbitals, which approximate the single-particle wave functions. 
This provides a description of the full quantum state (accounting for the spin-orbit coupling and phase), so we are able to account for all terms in the hyperfine Hamiltonian. 
%In addition, $\mathbf{k}\cdot\mathbf{p}$ corrections are included in the procedure to account for the inaccuracies of DFT (in calculating band gaps\cite{seidl1996generalized}). 

The hyperfine parameters for the conduction bands of GaAs and silicon have been established experimentally through measurements of the Knight shift.
The results found here from DFT for the conduction band of GaAs are consistent with Knight shift measurements in the fractional quantum Hall regime.\cite{khandelwal1998optically, desrat2013dispersive}
For silicon, the Knight shift has been measured in n-doped bulk samples.\cite{sundfors1964nuclear}  
Density functional theory (without $\mathbf{k}\cdot\mathbf{p}$) has been used to calculate the hyperfine constants,\cite{assali2011hyperfine} however the results are inconsistent with the Knight shift measurements of Ref.~\onlinecite{sundfors1964nuclear}. 
In contrast, we find here that a combined DFT+$\mathbf{k}\cdot\mathbf{p}$ procedure yields hyperfine constants for electrons in silicon that are consistent with the experiments of Ref.~\onlinecite{sundfors1964nuclear}.
We further apply this procedure to the valence-band (hole-spin) states of GaAs and silicon where we expect similarly accurate results.
There have been fewer experiments focused on the hole hyperfine interaction.
Experiments thus far have relied on extracting hole hyperfine couplings in GaAs (and InGaAs) quantum dots through the ratio of the Overhauser shifts for electrons and holes.\cite{fallahi2010measurement,  chekhovich2013element, vidal2016hyperfine, prechtel2016decoupling} 
Our theoretical results are roughly consistent with these ratios. 
Moreover, in silicon, we find hyperfine constants for holes that are consistent with recent $T_2^{\ast}$ times measured in silicon quantum dots,\cite{maurand2016cmos} suggesting those dephasing times may be limited by hyperfine interactions.

The remainder of this paper is organized as follows: 
In Sec.~\ref{sec:hyperfinenanostructures} we derive the hyperfine Hamiltonian in the envelope-function approximation accounting for relativistic effects (a finite Thomson radius) and write a projected effective hyperfine Hamiltonian for a nanostructure.
In Sec.~\ref{sec:hyperfineparameters} we define the hyperfine parameters for the states at the conduction-band minima and valence-band maxima of GaAs and silicon. 
In Sec.~\ref{sec:ElectronicStructure} we describe the procedure used to evaluate the hyperfine parameters, with the conclusions given in Sec.~\ref{sec:conclusions}.  Technical details are provided in Appendices A-F.

\section{Hyperfine interactions in nanostructures}\label{sec:hyperfinenanostructures}

The goal of this section is to parameterize the hyperfine interactions for a nanostructure in terms of parameters obtained from a bulk calculation.
This parameterization can be achieved within the envelope function approximation where the nanostructure confinement potential varies on a length scale that is large compared to the lattice constant of the host material. 
In nanostructures where the confinement has quickly-varying features on the scale of the lattice constant (e.g.~donors or acceptors in silicon with $1/r$ confining potentials),\cite{kohn1955hyperfine} the formalism developed here cannot be applied and other methods for calculating the hyperfine interactions become necessary.\cite{philippopoulos2019hole} 

The hyperfine interaction for a many-electron system in contact with nuclear spins $\bm{I}_l$ at sites $l$ in a nanostructure/molecule/etc.~can generally be written (setting $\hbar=1$) as 
\begin{equation}\label{eq:Hhf}
\mathcal{H}_{\mathrm{hf}}=\sum_l\gamma_{i_l}\bm{h}_l\cdot\bm{I}_l.
\end{equation}
Here, $\gamma_{i_l}$ is the gyromagnetic ratio of nuclear isotope $i_l$ at site $l$ and $\bm{h}_l$ is the hyperfine field operator acting on the many-electron spin/orbital space.  

We consider only non-magnetic semiconductors where spin polarization of the core electrons can be neglected.
In this case, finite contributions to the hyperfine field arise only from single-particle valence states associated with Bloch waves close to band extrema (valleys).  
We further assume a nanostructure defined by a slowly-varying potential that modulates a perfectly periodic crystal.
This is the regime of validity for the usual envelope-function approximation.
In this regime, we rewrite the hyperfine field in terms of a matrix $\mathbf{h}^j$ and a multicomponent field operator $\Psi(\mathbf{R})$. The matrix $\mathbf{h}^j$ depends only on the properties of the bulk crystal and atom $j$ (e.g. $j=\mathrm{Ga}, \mathrm{As}$ in GaAs) within the primitive cell, and $\Psi(\mathbf{R})$ accounts for the slowly-varying electronic spin/orbital/valley degrees of freedom, with $\mathbf{R}$ a lattice vector.
Further restricting to only short-range\footnote{In general, the hyperfine field at a nuclear site includes a contribution from electron density within typical atomic dimensions of the nuclear spin (short-range contribution) and a contribution from electron density localized at distant sites (long-range contribution).
The long-range contribution is suppressed by a factor $\sim (a_0/a)^3$, where $a_0$ is the Bohr radius and $a$ is a typical inter-atomic distance.
The long-range contribution is thus typically negligible (see Appendix C of Ref.~\onlinecite{fischer2008spin}).
However, in certain cases with large orbital currents or in materials with large $g$-factors, the long-range contributions can be significant (see Ref.~\onlinecite{yafet1961hyperfine}).} contributions to the hyperfine coupling leads to a local (contactlike) form,
\begin{equation}\label{eq:NuclearField}
\bm{h}_l \simeq v_0\Psi^\dagger(\mathbf{R}_l) \mathbf{h}^{j_l} \Psi(\mathbf{R}_l),
\end{equation}
where $v_0$ is the volume per atom [e.g. $v_0=\Omega/2$ for a primitive-cell volume $\Omega$ containing two atoms, as is the case for diamond (silicon) and zincblende (III-V) lattices considered below].  The matrix $\mathbf{h}^{j_l}$ depends only on the atomic species $j_l$ at site $l$ (and not the isotope $i_l$) provided we neglect the isotope mass effect,\cite{sekiguchi2014host} consistent with a Born-Oppenheimer approximation.  The vector $\mathbf{R}_l$ is the lattice vector that locates the primitive cell containing site $l$ (e.g., if $l$ and $l'$ are in the same primitive cell, then $\mathbf{R}_l = \mathbf{R}_{l'}$).
The multicomponent field operator $\Psi(\mathbf{R})$ has elements 
\begin{equation}
\Psi_\nu(\mathbf{R})=\frac{1}{\sqrt{V}}e^{i\mathbf{k}_\nu\cdot\mathbf{R}}\sum_\mathbf{q} e^{i\mathbf{q}\cdot\mathbf{R}} c_{\mathbf{q}\nu},
\end{equation}
with crystal volume $V$ and where $c_{\mathbf{q}\nu}$ annihilates an electron in an envelope state with band/valley index $\nu$, valley wavevector $\mathbf{k}_\nu$, and $\left|\mathbf{q}\right|$ is small compared to any reciprocal lattice vector.
The matrix $\mathbf{h}^j$ describes the short-range contributions to the hyperfine field for atom $j$ at position $\bm{\delta}_j$ within the primitive cell.
The associated matrix elements are 
\begin{eqnarray}
\mathbf{h}^{j}_{\nu\nu'}  & = & \int_\Omega d^3 r \psi^\dagger_\nu(\mathbf{r})\mathbf{h}(\mathbf{r}-\bm{\delta}_j)\psi_{\nu'}(\mathbf{r}),\label{eq:hMatrixElements}\\
\mathbf{h}(\mathbf{r}) & = & \frac{\mu_0}{4\pi}\left(2\mu_B\right)\left(\frac{\boldsymbol{\sigma}}{2}\cdot\overleftrightarrow{T}(\mathbf{r})+\sigma_0\frac{1}{r^3}f_\mathrm{T}(r)\mathbf{L}\right),\label{eq:hrSingleParticleMatrix}\\
f_\mathrm{T}(r) & = & \frac{r}{r+r_\mathrm{T}/2}\label{eq:fT},
\end{eqnarray} 
where $\psi_{\nu}(\mathbf{r})=e^{i\mathbf{k}_{\nu}\cdot\mathbf{r}}u_\nu(\mathbf{r})$.
Here, the spinor $u_\nu(\mathbf{r})=\left[u_\nu^\uparrow(\mathbf{r}),u_\nu^\downarrow(\mathbf{r})\right]^T$ describes the lattice-periodic Bloch amplitude for the Bloch wave at wavevector $\mathbf{k}=\mathbf{k}_\nu$.
We have chosen to normalize the Bloch amplitudes according to the convention:
\begin{equation}\label{eq:normBlock}
\int_\Omega d^3 r u_\nu^\dagger(\mathbf{r})u_{\nu'}(\mathbf{r})=\frac{\Omega}{v_0}\delta_{\nu \nu'}.
\end{equation}
In Eq.~\eqref{eq:hrSingleParticleMatrix}, $\mu_0$ is the vacuum permeability, $\mu_\mathrm{B}$ is the Bohr magneton, we have taken the bare electron $g$-factor to be $g\simeq 2$, $\boldsymbol{\sigma}$ is the vector of Pauli matrices, and $\sigma_0$ is the $2\times 2$ identity matrix.
The second term in Eq.~\eqref{eq:hrSingleParticleMatrix} describes coupling of the nuclear magnetic moment to the charge current generated by the electron angular momentum, $\mathbf{L}=\mathbf{r}\times(-i\pmb{\nabla})$.
The factor $f_\mathrm{T}(r)$ accounts for a cutoff at short distances on the order of the Thomson radius for a nucleus of charge $Z|e|$, $r_\mathrm{T}=Z\alpha^2 a_0$ [where $\alpha=(1/4\pi\epsilon_0)e^2/\hbar c\simeq 1/137$ is the fine-structure constant and $a_0=\hbar/(m_e c\alpha)$ is the Bohr radius].
The tensor $\overleftrightarrow{T}(\mathbf{r})$ accounts for both the Fermi-contact and magnetic dipole-dipole interactions, with tensor elements: 
\begin{eqnarray}\label{eq:tensor}
T^{\alpha\beta}(\mathbf{r})&=&\frac{8\pi}{3}\delta_\mathrm{T}(\mathbf{r})\delta_{\alpha\beta}+\frac{3 r_\alpha r_\beta-r^2\delta_{\alpha\beta}}{r^5}f_\mathrm{T}(r),\label{eq:Ttensor}\\
\delta_\mathrm{T}(\mathbf{r})&=&\frac{1}{4\pi r^2}\frac{df_\mathrm{T}(r)}{dr},\label{eq:deltaT}
\end{eqnarray}
where $\alpha,\beta \in\{x,y,z\}$.  Equation \eqref{eq:hrSingleParticleMatrix}, with \eqref{eq:Ttensor}, includes relativistic effects due to a finite Thomson radius $r_\mathrm{T}\ne 0$.  These relativistic effects can be significant for large-$Z$ atoms,\cite{breit1930possible,pyykko1973hydrogen,bluegel1987hyperfine,vanLeeuwen1994exact} so they are included here.

\begin{figure}
\centering
\includegraphics[width=\columnwidth]{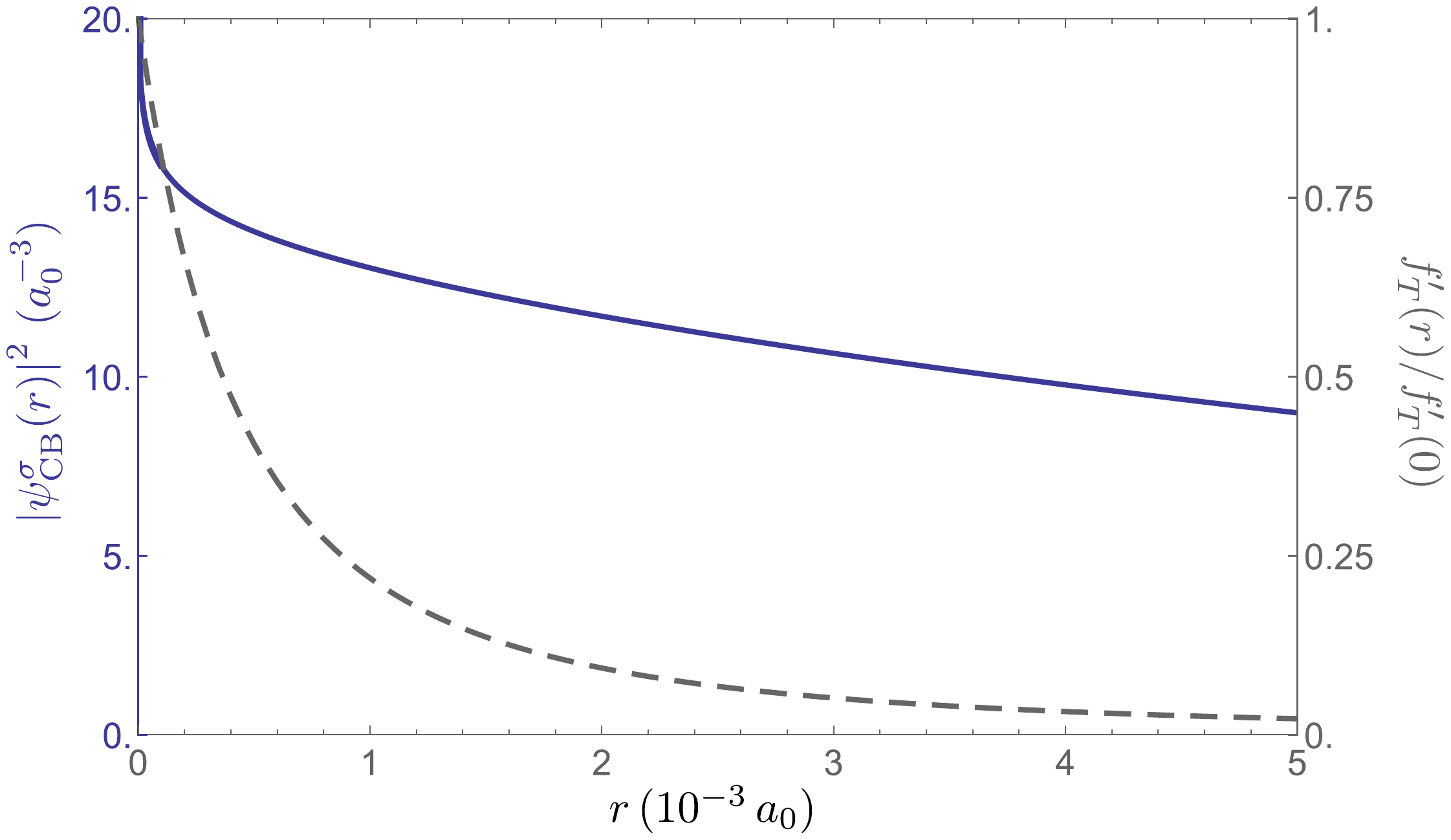}
\caption[Conduction-band density near As site]{Electron density near the As site in GaAs.
The density is found from the lowest unoccupied Kohn-Sham orbital in the conduction band at $\mathbf{k}=0$, $\psi^\sigma_\mathrm{CB}(r)$  (blue solid line, left axis).
The weight function $f'_\mathrm{T}(r)=4\pi r^2 \delta_\mathrm{T}(\mathbf{r})=(r_\mathrm{T}/2)/(r+r_\mathrm{T}/2)^2$ (gray dashed line, right axis) is used to evaluate the contact hyperfine coupling.
For As ($Z=33$), the Thomson radius is $r_\mathrm{T}=Z\alpha^2 a_0=1.76\times 10^{-3}\,a_0$.}\label{fig:CBDensity}
\end{figure}

Relativistic effects due to $r_\mathrm{T}\ne 0$ have been neglected in other approaches,\cite{vandewalle1993first, assali2011hyperfine} but we find that these corrections are essential for the present analysis.
In particular, our calculations make use of a basis of optimized single-particle states based on the scalar relativistic equation.\cite{vanLeeuwen1994exact,elk}
The $s$-like ($l=0$) solutions to the scalar relativistic equation show a weak (integrable) divergence close to a pointlike nucleus, necessitating the cutoff in Eq.~\eqref{eq:deltaT} (see also Fig.~\ref{fig:CBDensity}).  

An additional common simplification is to neglect the angular-momentum term in Eq.~\eqref{eq:hrSingleParticleMatrix} (see, e.g., Refs.~\onlinecite{vandewalle1993first, assali2011hyperfine}).
In this approach, the hyperfine couplings are expressed purely in terms of the electron spin density, without direct reference to the single-particle states and their associated phase information.
This procedure can be justified when calculating the isotropic Fermi contact term due to $s$-like states, but for states having a partial-wave expansion with $l\ne 0$ (as we consider below for the valence bands of silicon and GaAs), the angular-momentum term can give a significant contribution to the hyperfine coupling.   
For example, for a $p$-like heavy-hole state, $\ket{J=3/2, l=1, m_J =3/2}$ (where $J$ represents the total angular momentum, $l$ gives the orbital angular momentum, and $m_J$ is the angular momentum projected onto the relevant quantization axis), $\frac{\left|\left<\mathbf{L}\right>\right|}{\left|\left<\boldsymbol{\sigma}/2\right>\right|} = 2$, indicating that the nuclear orbital interaction represents a significant portion of the anisotropic hyperfine interaction [see Eq.~\eqref{eq:hrSingleParticleMatrix}] in this case.

\subsection{Effective Hamiltonian}
We take the hyperfine interaction to be weak compared to other electronic energy scales in a nanostructure, allowing us to consider a projected effective Hamiltonian.
When the electronic system can be well-described by a finite-dimensional quasi-degenerate subspace of low-energy states, $\{\ket{n}\}$,\footnote{Within the envelope function approximation, a general state $\ket{n}$ can be written as $\braket{\mathbf{r}}{n} =  \sum_{\nu}F^{n}_{\nu}(\mathbf{r}) \psi_{\nu}(\mathbf{r})$, where the sum, $\sum_{\nu}$ is over all spin/orbital/valley states and the $F^{n}_{\nu}(\mathbf{r})$ are the slowly-varying (on the scale of the lattice constant) envelope functions. 
In practice, the sum, $\sum_{\nu}$ is restricted to some reasonable set of orbitals and valleys.
For example, $\ket{n}$ can be taken to be an eigenstate of the Luttinger Hamiltonian, in which case the sum is restricted to the heavy-hole and light-hole states.} we consider the effective Hamiltonian, \begin{equation}\label{eq:effhf}
H_{\mathrm{hf}}=P\mathcal{H}_{\mathrm{hf}}P,
\end{equation}
where $P=\sum_{n}\ket{n}\bra{n}$ is a projector onto the finite-dimensional subspace $\{\ket{n}\}$.
The effect of the hyperfine interaction is then determined by the matrix elements $\bra{n}\bm{h}_l\ket{n'}$.

Equation \eqref{eq:effhf} applies to an arbitrary high-dimensional quasi-degenerate space, but a common case is when the ground space is only twofold degenerate.
For such a doubly degenerate ground space, $\{\ket{n}\}=\{\ket{+},\ket{-}\}$, Eq.~\eqref{eq:effhf} gives
\begin{equation}\label{eq:HhfTensor}
H_{\mathrm{hf}}=\sum_{l}\left[\bm{S}\cdot\overleftrightarrow{\mathbf{A}}_l\cdot\bm{I}_l+\gamma_{i_l}\mathbf{B}_l\cdot\bm{I}_l\right],
\end{equation}
where the hyperfine tensor $\overleftrightarrow{\mathbf{A}}_l$ and field $\mathbf{B}_l$ are given by:
\begin{eqnarray}
A_{l}^{\alpha\beta}  & = & 2\gamma_{i_l}\mathrm{Tr}\{S^{\alpha}h^{\beta}_l\},\label{eq:hftensor}\\
B^{\beta}_{l}  & = & \frac{1}{2}\mathrm{Tr}\{Ph^{\beta}_l\},\label{eq:hffield}
\end{eqnarray}
 and $\bm{S}$ is the vector of (pseudo)spin-1/2 operators:
\begin{eqnarray}
S^x & = & \frac{1}{2}\left(\left|+\right>\left<-\right|+\left|-\right>\left<+\right|\right),\label{eq:Sx}\\
S^y & = & \frac{1}{2i}\left(\left|+\right>\left<-\right|-\left|-\right>\left<+\right|\right),\label{eq:Sy}\\
S^z & = & \frac{1}{2}\left(\left|+\right>\left<+\right|-\left|-\right>\left<-\right|\right).\label{eq:Sz}
\end{eqnarray}
In the specific case where $\{\ket{+},\ket{-}\}$ form a Kramers doublet, related by time-reversal $\Theta$: $\Theta\ket{+}=e^{i\phi_0}\ket{-}$ (where $\phi_0$ is a global phase), then we have the further simplification $\mathbf{B}_{l}=\left(\bra{+}\bm{h}_l\ket{+}+\bra{-}\bm{h}_l\ket{-}\right)/2=0$.
This follows directly from the fact that $\bm{h}_l$ is odd under time reversal: $\Theta\bm{h}_l\Theta^{-1}=-\bm{h}_l$.
In this (common) scenario (as in the examples given below), the influence of the hyperfine interactions will be well-described by the hyperfine tensor matrix elements, $A_l^{\alpha\beta}$ alone.

\begin{table}
 \begin{ruledtabular}
\begin{tabular}{l|c|cc}
				& (electrons) 					&\multicolumn{2}{c}{(holes)}\\
  isotope ($i$) & $A^i$ ($\mu e\mathrm{V}$) & $A^i_{\parallel}$ ($\mu e\mathrm{V}$) & $A^i_{\perp}$ ($\mu e\mathrm{V}$) \\
   \hline
   $^{69}\mathrm{Ga}$ in GaAs & $74$ &  $1.4$ & $0.35$\\   
   $^{71}\mathrm{Ga}$ in GaAs & $94$ & $1.7$ & $0.45$ \\
   $^{75}\mathrm{As}$ in GaAs & $78$ & $11$ & $0.02$ \\ 
   $^{29}\mathrm{Si}$ in silicon & $-2.4$ &   $-2.5$ & $-0.01$  \\   
\end{tabular}\caption{Hyperfine parameters calculated for GaAs and crystalline silicon.
All parameters have been found from $\mathbf{k}=0$ Bloch amplitudes approximated by Kohn-Sham orbitals established in DFT using Elk, an all-electron DFT code\cite{elk} (see Sec.~\ref{sec:ElectronicStructure} for details).
    The silicon conduction-band parameter ($A^{^{29}\mathrm{Si}}$) is evaluated using DFT+$\mathbf{k}\cdot\mathbf{p}$ which accounts for the off-zone-center conduction-band minima in silicon.
The valence-band parameters ($A^i_{\parallel}$ and $A^i_{\perp}$) are given for a system where the isotope $i$ is located at an `$A$' site, with a neighboring (`$B$' site) atom at $\left(\frac{1}{4},\frac{1}{4},\frac{1}{4}\right)$ (see Sec.~\ref{sec:coordinates} and Fig.~\ref{fig:coordinates}). 
Numerical convergence has been verified for all parameters to within $2\%$ of the reported values.}\label{tab:hfparamsVB} 
 \end{ruledtabular}
 \end{table}

\subsection{Summary of key results}\label{sec:keyresults}

For a conduction-band electron confined to a nanostructure with a spin-independent envelope function, and for a fixed valley: $\Psi_\sigma(\mathbf{r})= F_\mathrm{e}(\mathbf{r})c_{\sigma}$, we can identify a two-level system $\ket{\pm}=c^{\dagger}_{\pm}\ket{0}$.
This allows us to apply Eq.~\eqref{eq:hftensor} with the spin operators given in Eqs.~\eqref{eq:Sx}, \eqref{eq:Sy}, \eqref{eq:Sz}.
If the electronic state is well-described by an $s$-like band, the isotropic contact interaction dominates, giving the well-known result for an electron spin in a quantum dot,\cite{merkulov2002electron,khaetskii2002electron}
\begin{equation}
A_l^{\alpha\beta} = A^{i_l}v_0\left|F_e(\mathbf{R}_l)\right|^2\delta_{\alpha\beta},
\end{equation}
where $A^{i_l}$ is the (bulk) contact hyperfine coupling for isotope $i_l$ at site $l$ (see Table \ref{tab:hfparamsVB}).

Alternatively, for the valence band of a zincblende III-V semiconductor (GaAs, InAs, InSb, etc.), or for the diamond-lattice form of a group IV element (Si, Ge, etc.), the states at $\mathbf{k}=0$ transform according to the $\Gamma_8$ irreducible representation of the $T_d$ double group.
For these states, we can project, for example, onto the two states that transform like states of angular momentum $J_z=m_J=\pm 3/2$: $\ket{\pm}=\ket{m_J=\pm 3/2}$ (the pure heavy-hole states).
These are separated in energy from the light-hole states ($\ket{m_J=\pm 1/2}$) under confinement or strain.
For these states, the $s$-wave component vanishes identically, and the dominant hyperfine coupling arises from the dipole-dipole and angular-momentum terms.
Assuming a pseudospin-independent envelope function for the heavy hole, $\Psi_{m_J}(\mathbf{r})=F_h(\mathbf{r})c_{m_J}$, Eq.~\eqref{eq:hftensor} gives
\begin{eqnarray}
A_l^{xx}&=&-A_l^{yy}= A^{i_l}_\perp v_0\left|F_h(\mathbf{R}_l)\right|^2,\label{eq:Aperp}\\
A_l^{zz}&=& A^{i_l}_\parallel v_0 \left|F_h(\mathbf{R}_l)\right|^2, \label{eq:Aparallel}
\end{eqnarray} 
where $A^i_{\parallel}$ and $A^i_{\perp}$ are valence-band hyperfine parameters (see Table \ref{tab:hfparamsVB}), and all other hyperfine-tensor elements vanish.
Here, the relation $A_l^{xx} \ne A_l^{yy}$ is a consequence of the fact that the diamond and zincblende lattices do not have a strict fourfold symmetry axis.
Equations \eqref{eq:Aperp} and \eqref{eq:Aparallel} apply in a coordinate system where the site $l$ is located at $(0,0,0)$ with a nearest-neighbor atom at $\left(\frac{1}{4},\frac{1}{4},\frac{1}{4}\right)$ in units of the cubic-cell lattice constant (see Sec.~\ref{sec:coordinates}).
More generally, the influence of the hyperfine coupling can be fully described in the four-dimensional subspace of heavy holes and light holes in terms of the same two coupling constants, $A^{i}_\perp$ and $A^{i}_\parallel$, given in Table \ref{tab:hfparamsVB} (see Sec.~\ref{sec:vbhyperfineparameters}). 

\begin{figure}
\centering
\includegraphics[width=\columnwidth]{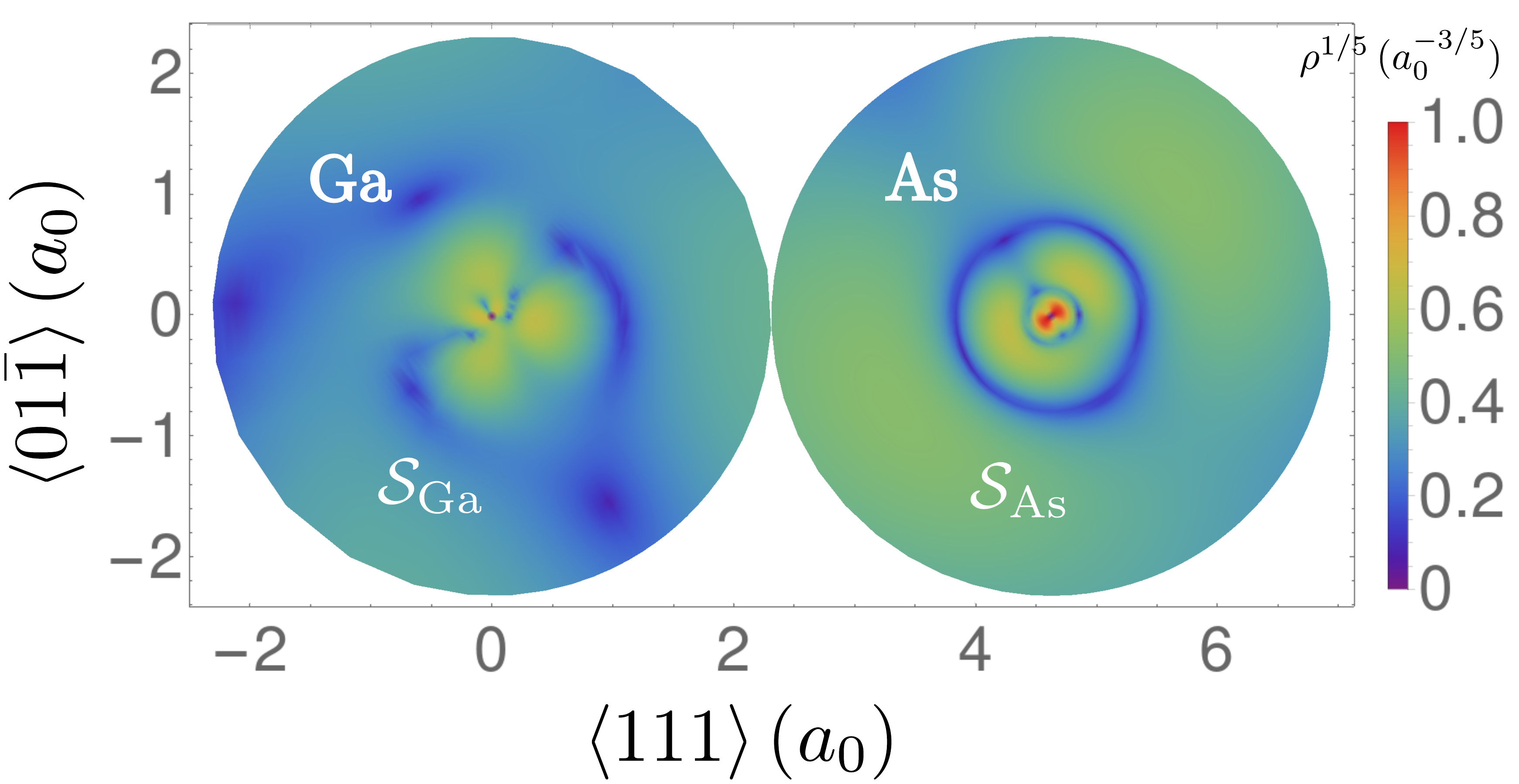}
\caption[2D density plot along (2,-1,-1) direction]{Calculated density, $\rho = \left|\phi_{3/2}(\mathbf{r})\right|^2$, of the $m_J=3/2$ heavy-hole state in GaAs, resulting from the Kohn-Sham orbital $\phi_{3/2}(\mathbf{r})$.
$\rho^{1/5}$ (instead of $\rho$) is plotted using a color scale (in units of $a_0^{-3/5}$, with $a_0$ the Bohr radius) so that the features of the density can be visible.
The density is shown along a cut in the $(2\bar{1}\bar{1})$ plane.
The spheres ${\cal{S}}_j$ define regions where the Kohn-Sham orbital $\phi_{3/2}(\mathbf{r})$ has been evaluated to extract the hyperfine tensor for atom $j=\mathrm{Ga,As}$.
Regions outside of the spheres ${\cal{S}}_j$ are shown in white.
The $p$ symmetry can be seen around the As sites (labeled), while the Ga site has a combination of $p$ and $d$ symmetry.
}\label{fig:fig2}
\end{figure}

The parameters $A^i$, $A^{i}_\perp$, and $A^{i}_\parallel$ fully characterize the bulk short-range hyperfine coupling for electrons in an $s$-like conduction band and for holes in a valence band that transforms according to the $\Gamma_8$ representation of the $T_d$ double group.
These parameters depend only on the isotope $i$, through the gyromagnetic ratio $\gamma_i$, and on the material-dependent microscopic Bloch functions $\psi_\nu(\mathbf{r})$ through the matrix elements given in Eq.~\eqref{eq:hMatrixElements}.
To approximately determine the relevant Bloch functions in GaAs and silicon, we have performed first-principles DFT calculations.
The $\mathbf{k}=0$ Bloch functions are then approximated directly with optimized Kohn-Sham orbitals (rather than the density alone), providing an accurate representation of the electron/hole states in the vicinity of atoms in the crystal (see Figs.~\ref{fig:CBDensity} and \ref{fig:fig2} for examples in the conduction and valence bands of GaAs, respectively).
To find accurate Bloch functions at an off-zone-center band extremum $\mathbf{k}=\mathbf{k}_\nu\ne 0$ (as is the case in the conduction band of silicon), we find it is necessary to determine the correct linear combination of $\mathbf{k}=0$ Kohn-Sham orbitals by diagonalizing an appropriate $\mathbf{k}\cdot\mathbf{p}$ Hamiltonian at $\mathbf{k}=\mathbf{k}_\nu$.
In each case, the integral in Eq.~\eqref{eq:hMatrixElements} is then evaluated numerically giving the hyperfine parameters.
The results are shown in Table \ref{tab:hfparamsVB} for GaAs and silicon.

For the conduction bands of GaAs and silicon (electrons), we find contact hyperfine couplings $A^i$ that are consistent with known experimental values (see Table \ref{tab:eta}). 
There have been fewer experimental studies related to the hyperfine coupling for holes. 
Moreover, in some cases, experiments on hole spins have led to conflicting interpretations.
On one hand, it has been argued that the hyperfine interaction in the heavy-hole subspace is predominantly Ising ($A^i_{\perp} \simeq 0$) because heavy-hole spin relaxation times have been measured to be consistent with a negligible transverse hyperfine coupling in self-assembled InGaAs quantum dots.\cite{gerardot2008optical} In addition, the heavy-hole transverse Overhauser shift has been observed to be small (again, in self-assembled InGaAs quantum dots).\cite{prechtel2016decoupling}
Measurements of tunneling between spin-resolved Landau levels in a two-dimensional hole gas in GaAs are also consistent with a negligible transverse hyperfine coupling.\cite{klochan2015landau} On the other hand, separate experiments measuring the longitudinal Overhauser shift in GaAs/AlGaAs and InGaAs/GaAs quantum dots have been interpreted to indicate a substantial $p$-$d$ hybridization of the valence-band states near the Ga sites, leading to non-negligible transverse hyperfine coupling to the Ga isotopes ($A^i_{\perp} \sim A^i_{\parallel}$). \cite{chekhovich2013element}
The results of this experiment, combined with the interpretation of Ref.~\onlinecite{chekhovich2013element}, also suggest substantial in-plane components of the total heavy-hole Overhauser field.
Because the DFT procedure used here gives direct access to the wave function, both the hyperfine couplings and the $p$-$d$ hybridization can be calculated (see Appendix~\ref{app:pd}). 
Here, we find an intriguing mix of the two descriptions: For heavy holes, the coupling to the As site is stronger and almost purely Ising (small transverse coupling), while the transverse coupling to the Ga site is a significant fraction of its longitudinal coupling (see Table~\ref{tab:hfparamsVB} and Appendix \ref{app:pd} for a possible explanation).
However, due to the larger (Ising) hyperfine coupling to the As nuclear spins, the total Overhauser field experienced by a heavy hole in a GaAs quantum dot will be oriented predominantly along the growth direction of the quantum dot, even for a randomly polarized nuclear-spin ensemble.
 The different behavior at Ga and As sites can be understood as follows: 
Because As is more electronegative than Ga, the hole is more highly localized around the As site in GaAs (see Fig.~\ref{fig:fig2}).
The potential experienced by the hole in the vicinity of the As site can thus be taken to be more spherically symmetric.
Therefore, close to the As site the hole wave function will approximate a pure angular-momentum eigenstate, a $p$ state (see Table~\ref{tab:pdhyb} in Appendix \ref{app:pd}, below). 
In contrast, the hole is not sufficiently tightly bound to the Ga atom to fully mask the potential due to neighboring As atoms.
In the vicinity of the Ga atom, the hole adapts to the reduced tetrahedral symmetry of the crystal and is therefore not in an angular-momentum eigenstate.
Instead, the hole wave function describes a $p$-$d$ hybridized state (see Fig.~\ref{fig:fig2} and Table~\ref{tab:pdhyb}).
This $p$-$d$ hybridization leads to non-Ising corrections to the heavy-hole hyperfine Hamiltonian. 
At the same time, the more delocalized nature of the hole wave function at the Ga sites leads to a significantly smaller hyperfine coupling (due to the larger average distance from the nucleus). 

%This explanation is supported by the calculated $p$-$d$ hybridization in the vicinty of the Ga and As site in GaAs (see Table \ref{tab:pdhyb})
%However, due to the larger (Ising) hyperfine coupling to the As nuclear spins, the total Overhauser field experienced by a heavy hole in a GaAs quantum dot will be oriented predominantly along the growth direction of the quantum dot, even for a randomly polarized nuclear-spin ensemble.

For holes in silicon, we find an Ising hyperfine coupling ($A^{^{29}\mathrm{Si}}_{\perp} \simeq 0$, see Table~\ref{tab:hfparamsVB}).
The strength of the coupling is comparable to the contact interaction ($A^{^{29}\mathrm{Si}}$) for the conduction band of silicon.
Typically, the anisotropic hyperfine coupling for ($p$-type valence-band) holes is assumed to be weaker (by a factor of $\sim 5$-$10$) than the contact hyperfine coupling for ($s$-type conduction-band) electrons.\cite{fischer2008spin} 
However, in silicon the states at the conduction-band minima are $s$-$p$ hybridized, reducing the effect of the contact interaction for conduction-band states. 
We find that this reduction leads to a value that is comparable to the (normally smaller) anisotropic hyperfine coupling in the valence band.

\section{Bulk hyperfine parameters}\label{sec:hyperfineparameters}

Any Bloch wave $\psi_{\nu}^{\sigma}(\textbf{r})$ can be described near an atomic site $j$ using the partial-wave expansion
\begin{equation}\label{eq:Blochexpansion}
\psi_{\nu}^{\sigma}(\textbf{r}+\bm{\delta}_{j})=\sum_{lm}R_{lm\sigma}^{j\nu}(r)Y_{lm}(\theta,\phi),
\end{equation}
where $R_{lm\sigma}^{j\nu}(r)$ are radial functions and $Y_{lm}(\theta,\phi)$ are the spherical harmonics. States that have a contribution entirely from the $l=0$ term to the sum in Eq.~\eqref{eq:Blochexpansion}, namely $s$-like states, are isotropic. Therefore, they have a vanishing dipolar and angular-momentum contribution to the hyperfine interaction and contribute only via the contact part of the hyperfine interaction, $\propto \delta_{\mathrm{T}}(\mathbf{r})$ [see Eq.~\eqref{eq:tensor}]. 

\begin{table}
 \begin{ruledtabular}
\begin{tabular}{llccc}
& & \multicolumn{2}{c}{GaAs} & silicon \\
& & $\eta^{\mathrm{Ga}}$  & $\eta^{\mathrm{As}}$ & $\eta^{\mathrm{Si}}$ \\
 \hline
(i)& DFT at $\mathbf{k}=0$ (+$\mathbf{k}\cdot\mathbf{p}$) & 2500 & 3800 & 88 \\
(ii)& Knight shifts (Refs.~[\onlinecite{khandelwal1998optically}], [\onlinecite{desrat2013dispersive}])& $2200$  & $3500$ &  $-$\\
(iii)& Estimates (Ref.~[\onlinecite{paget1977low}]) & 2600 & 4400 &  $-$\\
(iv)& Knight shift (Ref.~[\onlinecite{sundfors1964nuclear}]) & $-$ & $-$ & $100 \pm 10$ \\ 
(v)& DFT at $\mathbf{k}=\mathbf{k}_\nu$ (Ref.~[\onlinecite{assali2011hyperfine}]) & $-$ & $-$ & $159.4 \pm 4.5$ \\
 
\end{tabular} \caption{The parameter $\eta^j$ characterizing the degree of localization of an electron around atom $j$ [see Eq.~\eqref{eq:eta}].  This parameter, together with the gyromagnetic ratio $\gamma_i$, determines the contact hyperfine coupling for isotope $i$, $A^i$ [see Eq.~\eqref{eq:contacthfparameter}].
(i): Theoretical results from the present work. (ii): Experimental Knight shifts measured for spin-polarized electronic states in GaAs quantum wells have been used to extract $\eta^j$ using the procedure descibed in Appendix~\ref{app:Knight}.  (iii): Theoretical estimates reported in Paget et al.~(Ref.~\onlinecite{paget1977low}), extrapolated from measurements in InSb.  (iv): Experimental value of $\eta^j$ extracted from Knight-shift measurements in bulk silicon (Ref.~\onlinecite{sundfors1964nuclear}).  The error bar describes the standard deviation of the results of different measurements. (v): Theoretical value calculated by Assali et al., Ref.~\onlinecite{assali2011hyperfine}.  The error bar is based on a statistical error from different runs (with different supercell sizes).
}\label{tab:eta}
 \end{ruledtabular}
 \end{table}

\subsection{Conduction bands with $s$-like Bloch functions}\label{sec:cbhfparams}
 
We consider states coming from different equivalent valleys of an $s$-like band, such as the conduction-band states of GaAs (1 valley) and silicon (6 valleys). In the limit of weak spin-orbit coupling, these states can be written as product states of spin and orbit, which means that the index $\nu=(v,\chi)$, where $v$ labels the orbital (valley) and $\chi$ labels the spin so that the Bloch amplitudes can be written as  $\psi_{v,\chi}^{\sigma}(\mathbf{r})=\psi_{v,\sigma}^{\sigma}(\mathbf{r})\delta_{\sigma \chi}$. We further assume that the valleys are related by space-group transformations of the crystal so that 
\begin{equation}
R_{00\sigma}^{jv\sigma}(r) = R^j_s(r)\quad\forall\,v,\sigma,
\end{equation}
i.e. the radial function associated with the $s$ part of the Bloch function is identical for all valleys. For these $s$ states, the matrix elements of $\mathbf{h}^j$ are given by:
\begin{equation}
\mathbf{h}^j_{v\chi v'\chi'}=\frac{2\mu_0}{3}\mu_B\frac{\eta^j}{v_0}\bm{\sigma}_{\chi \chi'},
\end{equation}
where the dimensionless parameter\cite{shulman1956nuclear,sundfors1964nuclear} 
\begin{equation}\label{eq:eta}
\eta^j=v_0 \left<\left|R^j_s(r)\right|^2\right>_{\delta_{\mathrm{T}}}
\end{equation} 
characterizes the degree of localization of the electron at the atom $j$ and is independent of the valley index $v$ because we have assumed that all valleys are equivalent (see Table \ref{tab:eta}). In Eq.~\eqref{eq:eta}, we have introduced the notation 
\begin{equation}\label{eq:defav}
\left< f(r)\right>_{g}=\int_0^{\infty} f(r) g(r)r^2dr
\end{equation}
to indicate a weighted average of the function $f$ with respect to the weighting function $g$. The contact part of the hyperfine Hamiltonian can also be characterized by the parameter
\begin{equation}\label{eq:contacthfparameter}
A^{i}=\frac{4\mu_0}{3}\mu_B\gamma_{i}\frac{\eta^{j_i}}{v_0},
\end{equation}
where $j_i$ labels the atom associated with isotope $i$ (see Table \ref{tab:hfparamsVB}).

\subsection{Valence-band holes}\label{sec:vbhyperfineparameters}

We consider here a subspace spanned by states that transform according to the $\Gamma_8$ representation of the $T_d$ double group.
Examples include the states at the valence-band maxima of silicon and III-V semiconductors such as GaAs. 

A simple basis for the $\Gamma_8$ representation of the $T_d$ double group is composed of the four states with total angular momentum $J=3/2$ and orbital angular momentum $l=1$.
Without loss of generality, we take the $[001]$ direction (the $z$-axis) to be a relevant quantization axis.
Under this convention, the states that transform like the states with $m_J=\pm3/2$ units of angular momentum about $\hat{z}$ are the heavy-hole states and those that transform like the $m_J=\pm1/2$ states are light-hole states.
In this four-dimensional subspace, we can therefore label the states with the allowed $m_J$ values, so that $\nu \in \{-3/2,-1/2,1/2,3/2\}$. %, in spite of the fact that other angular momentum states can contribute to their expansion. %
If the expansion from Eq.~\eqref{eq:Blochexpansion} is performed up to $l=2$ for each state (see Appendix \ref{app:GTPO}), the four Bloch amplitudes at the valence-band maximum can be parametrized by three different real radial functions, 
\begin{align}\label{eq:radfncs}
& R_p^{j}(r)=R_{1,1,\uparrow}^{j,3/2}(r),\\\label{eq:radfncs2}
& R_d^{j}(r)= i R_{2,-1,\uparrow}^{j,3/2}(r),\\\label{eq:radfncs3}
& R_{d'}^{j}(r)=i R_{2,0,\downarrow}^{j,3/2}(r).
\end{align}
The remaining radial functions $R_{l,m,\sigma}^{j,m_J}(r)$ either vanish or are linear combinations of these three (see Appendix \ref{app:GTPO}).
Even though the $d'$ orbital is allowed by symmetry, it is often neglected, even in works where $d$-orbital hybridization for the hole states is taken into account.\cite{chekhovich2013element}
Because this orbital corresponds to a state with opposite spin [$\downarrow$, in this case, Eq.~\eqref{eq:radfncs3}] relative to the $p$ and $d$ orbitals in the wave function [$\uparrow$, Eqs.~\eqref{eq:radfncs} and~\eqref{eq:radfncs2}], we expect the weight of the $d'$ orbital, or equivalently the magnitude of the $R_{d'}^{j}(r)$ radial function, to be more significant in materials with large spin-orbit coupling.

In the subspace of heavy holes and light holes, the matrix $\mathbf{h}^j$, given by Eqs.~\eqref{eq:hMatrixElements}, \eqref{eq:hrSingleParticleMatrix}, and \eqref{eq:fT}, can be expressed as a linear combination of the angular-momentum matrices for a spin-$3/2$ particle, $\mathrm{J}_{\beta}$, and $\mathrm{J}_{\beta}^3$, ${\beta} \in \{x,y,z\}$ \cite{luttinger1956quantum,vidal2016hyperfine}
\begin{equation}\label{eq:hjVB}
\mathbf{h}^j=\left(\frac{1}{3}h^{j}_{\parallel}-\frac{3}{2}h^{j}_{\perp}\right) \textbf{J} + \frac{2}{3}h^{j}_{\perp} {\cal{J}},
\end{equation}
where ${\cal{J}} = (\mathrm{J}_x^3,\mathrm{J}_y^3,\mathrm{J}_z^3)$, and where $h^{j}_{\perp}$ and $h^{j}_{\parallel}$ are two hyperfine parameters.
These two parameters  can be written in terms of the matrix elements of $1/r^3$ as
\begin{align}\label{eq:ahfparams1}
h^{j}_{\parallel}=\frac{\mu_0}{2\pi} \mu_B&\left[\frac{8}{5}M^{j}_{p,p}-\frac{12}{7}M^{j}_{d,d}-\frac{4}{7}M^{j}_{d',d'}\right.\nonumber\\
&\left.+\frac{4}{7}\sqrt{\frac{3}{2}}\mathrm{Re}\left(M^{j}_{d,d'}\right)\right],
\end{align}
\begin{equation}\label{eq:ahfparams2}
h^{j}_{\perp}=\frac{\mu_0}{2\pi}\mu_B\left[\frac{6}{7}M^{j}_{d,d}+\frac{2}{7}M^{j}_{d',d'}-\frac{30}{7}\sqrt{\frac{3}{2}}\mathrm{Re}\left(M^{j}_{d,d'}\right)\right],\nonumber
\end{equation}
where 
\begin{equation}\label{eq:M}
M^{j}_{\lambda\lambda'}=\left<\frac{R_{\lambda}^{j}(r)R^{j}_{\lambda'}(r)}{r^3}\right>_{f_{\mathrm{T}}},
\end{equation}
for $\lambda,\lambda'\in\{p,d,d'\}$, the numerical factors arise from angular integrals, and $f_{\mathrm{T}}$ is the weighting function given by Eq.~\eqref{eq:fT}.
These two parameters can also be expressed in units of energy (see Table \ref{tab:hfparamsVB}) as 
\begin{equation}
A_{\perp / \parallel}^i=\gamma_i h^{j_i}_{\perp / \parallel}.
\end{equation}

\subsection{Choice of coordinate system}\label{sec:coordinates}

\begin{figure}
\centering
\includegraphics[width=\columnwidth]{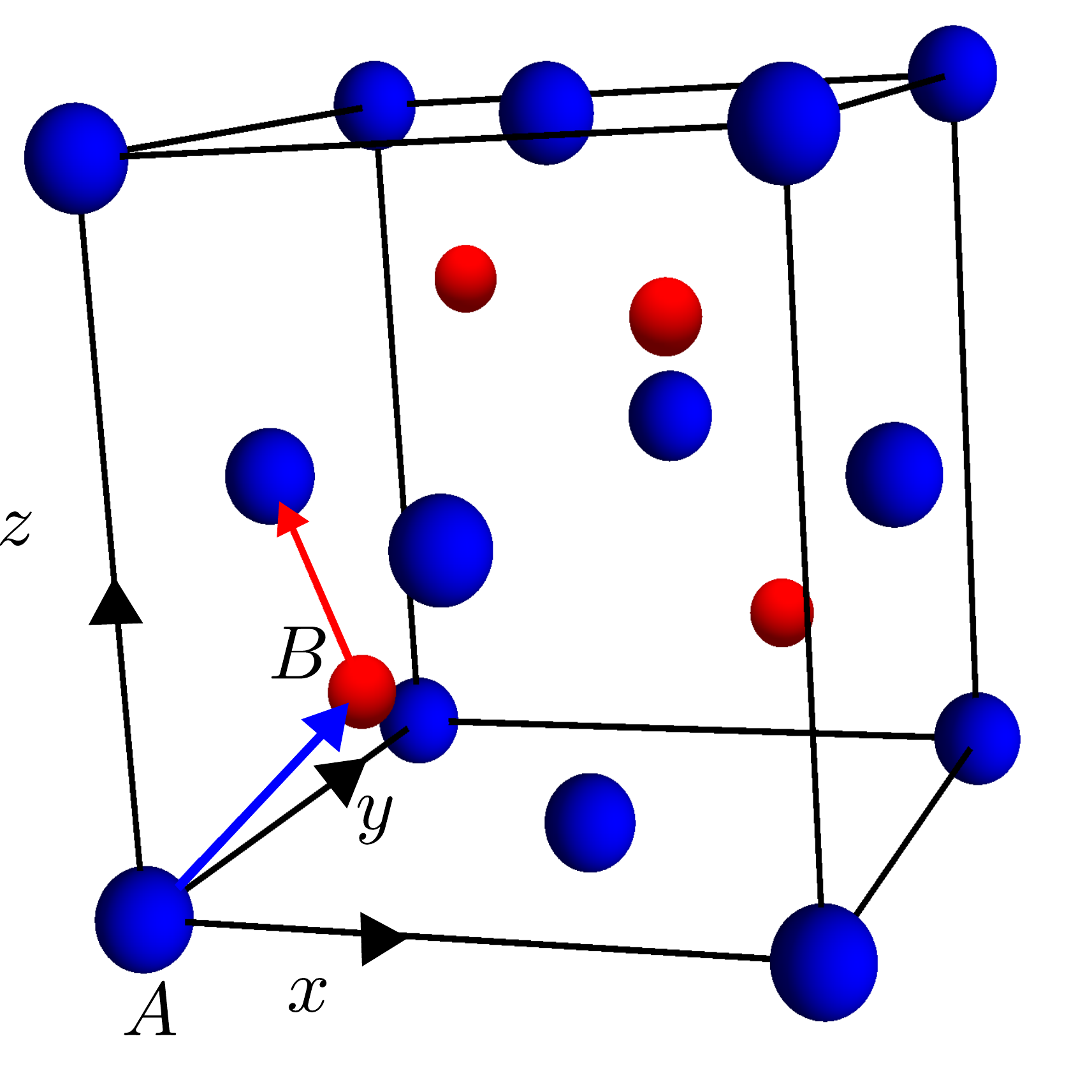}
\caption{Cubic unit cell for a zincblende or diamond lattice.
The blue and red spheres represent the two inequivalent sites in the zincblende lattice.
We have chosen the blue atom to be at the origin, $0 0 0$ ($A$ site, see main text), and a red atom to be located at $\frac{1}{4}\frac{1}{4}\frac{1}{4}$ ($B$ site).
An $A$ site can be related to a $B$ site by performing a translation of the coordinate system by $(\frac{1}{4}, \frac{1}{4}, \frac{1}{4})$, represented by the blue arrow, followed by a rotation of the coordinate system by $\pi/2$ about the $z$ axis.
The red arrow represents the vector $(\frac{1}{4}, \frac{1}{4}, \frac{1}{4})$ in the coordinate system with the $B$ site at the origin, $000$, and an $A$ site at $\frac{1}{4}\frac{1}{4}\frac{1}{4}$. }\label{fig:coordinates}
\end{figure}

Crystals break pure rotational symmetry, therefore their electronic eigenstates cannot in general be written as pure angular-momentum eigenstates.
For example, in the valence bands of GaAs and silicon, the eigensates can be approximated by a linear combination of $p$ and $d$ orbitals (see Sec.~\ref{sec:vbhyperfineparameters}).
The presence of the $d$ orbitals reflects the tetrahedral symmetry of the crystal and introduces the term proportional to ${\cal{J}}$ in the matrix $\mathbf{h}^j$ [Eq.~\eqref{eq:hjVB}] which has consequences on the symmetries of the hyperfine tensor. 

In both GaAs and silicon, the coordinate system can be set up so that the cubic unit cell has one nucleus at $0 0 0$, and another nucleus at $\frac{1}{4} \frac{1}{4} \frac{1}{4}$ (Fig.~\ref{fig:coordinates}).
Given this specific coordinate system, we can label as $A$ all the sites related to $000$ by a lattice vector and all sites related to $\frac{1}{4} \frac{1}{4} \frac{1}{4}$ by a lattice vector are labeled by $B$, with the understanding that all $A$ sites are equivalent and all $B$ sites are equivalent.

Equations \eqref{eq:Hhf} and \eqref{eq:NuclearField} for the hyperfine Hamiltonian within the envelope-function approximation can be combined to define a Hamiltonian matrix associated with site $l$, $\mathrm{H}^l$:
\begin{equation}
{\cal{H}}_{\mathrm{hf}} = \sum_l v_0 \Psi^\dagger(\mathbf{R}_l)  \mathrm{H}^{l} \Psi(\mathbf{R}_l),
\end{equation}
with 
\begin{equation}\label{eq:hfHamnoEnv}
\mathrm{H}^{l} = \gamma_{i_l} \mathbf{h}^{j_l} \cdot \bm{I}_l.
\end{equation}
Here, we recall that $i_l$ indicates the isotope situated at site $l$ and $j_l$ indicates the atom situated at site $l$. 
The hyperfine matrix $\mathrm{H}^{l}$ is simply the hyperfine Hamiltonian matrix expressed in the basis of Bloch states, $\psi_{\nu}(\mathbf{r})$ [see Eq.~\eqref{eq:hMatrixElements}]. 
In the subspace of valence-band states, this matrix is given by inserting $\mathbf{h}^{j}$ from Eq.~\eqref{eq:hjVB} into Eq.~\eqref{eq:hfHamnoEnv}.
Restricting further to the heavy-hole subspace, and for an isotope $i_l$ located at an $A$ site labeled by $l$, the matrix is
\begin{equation}\label{eq:hfA}
\mathrm{H}^{l,A}_{\mathrm{HH}}=\frac{1}{2}\left[A^{i_l}_{\parallel}\sigma_zI^{l,A}_z+A^{i_l}_{\perp}\left(\sigma_xI^{l,A}_x-\sigma_yI^{l,A}_y\right)\right],
\end{equation}
where $I^{l,A}_{\alpha}$ are the nuclear spin operators for the nuclear spin at the $A$ site labeled by $l$ and $\sigma_{\alpha}$ are Pauli matrices. As can be seen in Fig.~\ref{fig:coordinates}, an $A$ site can be related to a
$B$ site by performing a translation of the coordinate system by $( 1/4 , 1/4 , 1/4 )$ followed
by a rotation of the coordinate system by $\pi/2$ about the $z$ axis.
The result of this rotation is that $x \rightarrow y$ and $y \rightarrow -x$.
Under this rotation, $A^{xx} \rightarrow A^{yy}$ and $A^{yy} \rightarrow A^{xx}$.
Therefore, in the same coordinates used to describe $\mathrm{H}^{l,A}_{\mathrm{HH}}$, the hyperfine coupling for an isotope $i_{l'}$ located at a $B$ site ($l'$) is 
\begin{equation}\label{eq:hfB}
\mathrm{H}^{l',B}_{\mathrm{HH}}=\frac{1}{2}\left[A^{i_{l'}}_{\parallel}\sigma_zI^{l',B}_z - A^{i_{l'}}_{\perp}\left(\sigma_xI^{l',B}_x-\sigma_yI^{l',B}_y\right)\right],
\end{equation}
which has the opposite sign for the term with coefficient $A_{\perp}$ relative to the Hamiltonian for the $A$ sites, $\mathrm{H}^{l,A}_{\mathrm{HH}}$ [see Eq.~\eqref{eq:hfA}]. 
This sign difference for $A$ sites and $B$ sites may lead to non-trivial interference effects in the dynamics of hole spins confined to III-V and group IV nanostructures. 
We give the valence-band hyperfine couplings for GaAs and silicon in Table \ref{tab:hfparamsVB}.
In each case, the couplings are given assuming a coordinate system where the isotope in question is at an $A$ site.  
We also present the light-hole hyperfine Hamiltonian matrix in Appendix \ref{app:LHHam}.

\section{First-principles electronic structure}\label{sec:ElectronicStructure}

In Eq.~\eqref{eq:NuclearField}, the multi-component field operator, $\Psi(\bf{R})$, acts on the envelope functions, while $\mathbf{h}^j$ accounts for the short-range electronic structure, determined by the Bloch waves, $\psi_{\nu}(\mathbf{r})$.
The matrix $\mathbf{h}^j$ can thus be found from a bulk calculation for the translationally-invariant crystal.  Here, we calculate $\mathbf{h}^j$ using DFT. 

%We have used DFT to evaluate the hyperfine parameters for the conduction and valence bands of GaAs and silicon.
Hyperfine parameters are often evaluated through the density alone.
\cite{feller1984ab, vandewalle1993first, yazyev2008hyperfine, gali2008ab, assali2011hyperfine,ghosh2019all}  
Because the matrix elements of orbital angular momentum, $\mathbf{L} = \mathbf{r} \times (-i \boldsymbol{\nabla})$, depend on the phase of the wave function, the contribution from the nuclear-orbital interaction ($\sim \mathbf{L} \cdot \bm{I}$) to the hyperfine parameters $h_{\parallel}^j$ and $h_{\perp}^j$ cannot generally be calculated using the density alone.
This contribution is therefore often neglected.\cite{vandewalle1993first, assali2011hyperfine,ghosh2019all}
Here, we assume the Kohn-Sham orbitals, $\phi_{\nu}(\bf{r})$, can approximate the Bloch waves, $\psi_{\nu}(\mathbf{r})$ (as has been done, e.g., in Ref.~\onlinecite{wu2016spin}).
This approximation is valid at least when correlations are weak, so that the many-body ground state is well described by a single Slater determinant (Hartree-Fock limit).  

All of the DFT calculations presented here are done using the Elk code \cite{elk} with the exchange-correlation functional of Perdew, Burke and Ernzerhof (GGA-PBE).\cite{perdew1996generalized}
Elk is an all-electron code that avoids potential pitfalls associated with extracting the short-range electronic structure from a pseudopotential.\cite{assali2011hyperfine}
Within Elk, the Kohn-Sham orbitals for the valence electrons are calculated by solving the Dirac equation under the scalar relativistic approximation,\cite{takeda1978scalar} so it is essential to use the relativistic form of the hyperfine interaction to find accurate results. 

To compute the hyperfine parameters we run Elk (with input file set for ``very high quality" [vhq parameter] convergence),\cite{elk} to compute the Kohn-Sham orbitals at the conduction-band minima and valence-band maxima of GaAs and silicon.
We then treat these Kohn-Sham orbitals as approximations for the Bloch waves, $\psi_{\nu}(\mathbf{r}) \approx \phi_{\nu}(\mathbf{r})$. 

\subsection{Conduction band of GaAs }\label{sec:numevalcbgaas}

The Kohn-Sham orbital at the conduction-band minimum of GaAs ($\mathbf{k} = 0$) is found to be almost completely $s$-like (see Appendix \ref{app:snorm}).
As explained in Sec.~\ref{sec:cbhfparams}, this symmetry property of the wave function implies that the hyperfine interaction will be dominated by the contact term.
The integral for the contact hyperfine interaction has a weighting function, $\delta_{\mathrm{T}}(r)$, that weights the points within a distance $r_{\mathrm{T}}^j$ from the nuclei strongly, where $r_{\mathrm{T}}^j$ is the Thomson radius for atom $j$.
It is therefore important to find an accurate description of the Kohn-Sham orbital at short length scales ($r\lesssim r_{\mathrm{T}}^j$).
We sample the wave function on an equally spaced one-dimensional (radial) grid of points starting from each atom $j$ (Ga or As) within the unit cell out to a distance of $100 r_{\mathrm{T}}^j$.
These values represent a numerical description of the conduction-band wavefuntion $\psi(\mathbf{r}+\boldsymbol{\delta}_j)$.
Because the $s$-component of the wave function is spherically symmmetric, the radial functions are easily determined using $\psi(\mathbf{r}+\boldsymbol{\delta}_j)=Y_0^0(\theta, \phi) R_s^j(r) = R_s^j(r)/\sqrt{4\pi}$.
Once the radial functions have been obtained, we numerically evaluate the integral from Eq.~\eqref{eq:eta} (see Appendix \ref{app:snorm}).  

Once the integral $\left<\left|R^j_s(r)\right|^2\right>_{\delta_{\mathrm{T}}}$ has been evaluated, it can be used with Eqs.~\eqref{eq:eta} and \eqref{eq:contacthfparameter} to evaluate $\eta^j$ and the contact hyperfine parameter $A^i$ for the isotopes of Ga and As in GaAs.
We have verified that $\eta^j$ has converged with respect to certain parameters (e.g., the number of basis states and the density of $k$-points for which the calculation is performed; see Appendix \ref{app:conv} for the full list) to within 1\%  of its asymptotic value (see Appendix \ref{app:conv} for details).
The resulting hyperfine constants (given in Table \ref{tab:hfparamsVB}) are consistent with the accepted values estimated by Paget et al.~(Ref.~\onlinecite{paget1977low}).
The accuracy of this estimate may be in question since it is based on measurements in an analogous material (InSb), rather than direct measurements in GaAs.
However, the hyperfine constants calculated here are also consistent with Knight-shift measurements made on (fractional and integer) quantum-Hall states in GaAs quantum wells\cite{khandelwal1998optically,desrat2013dispersive} (see Table~\ref{tab:eta}, and Appendix~\ref{app:Knight} for details). 

\subsection{Conduction band of silicon }\label{sec:numevalcbsi}
In contrast to GaAs, the conduction band of silicon has six minima (valleys).
Each minimum is situated at roughly $84 \%$ of the way to any of the six equivalent $X$ points from the $\Gamma$ point.
The states at the conduction-band minima of silicon are $s$-$p$ hybridized.
Even though an anisotropic hyperfine interaction is not forbidden by symmetry (due to the $s$-$p$ hybridization), previous theoretical studies indicate that the contact part of the hyperfine Hamiltonian dominates over the anisotropic piece in bulk silicon.\cite{assali2011hyperfine}
We therefore neglect the anisotropic hyperfine interaction when investigating the hyperfine coupling in the conduction band of silicon. Because only $s$ states have a non-vanishing contact hyperfine interaction, we project onto the $s$-like component of the states at the conduction-band minima and use the same method described in Sec.~\ref{sec:numevalcbgaas} to evaluate the hyperfine constants for these states.
The result, $\eta^{\mathrm{Si}} = 160$, is consistent with  the theoretical result of Ref.~\onlinecite{assali2011hyperfine}, $\eta^{\mathrm{Si}}=159.4\pm4.5$, which was obtained using the Wien2k\cite{WIEN2K} all-electron DFT code with the non-relativistic formula for the contact-hyperfine constant [taking the limit as $r^j_{\mathrm{T}}\rightarrow 0$ in Eq.~\eqref{eq:Ttensor}]. 
Both of these calculations for the density directly at the conduction-band minima are, however, inconsistent with the measured value $\eta^{\mathrm{Si}}=100 \pm 10$, reported in Ref.~\onlinecite{sundfors1964nuclear}, obtained from Knight-shift and Korringa-relaxation measurements.\footnote{The authors of Ref.~\onlinecite{assali2011hyperfine} compared their result, $\eta^{\mathrm{Si}}=159.4\pm4.5$, with the result reported in Ref.~\onlinecite{shulman1956nuclear}, $\eta^{\mathrm{Si}} = 186 \pm 18$, which was derived from $^{29}\mathrm{Si}$ nuclear-spin relaxation measurements.
Although these results seem roughly consistent with each other, there was an error found in the analysis of  Ref.~\onlinecite{shulman1956nuclear} which, when accounted for, leads to $\eta^{\mathrm{Si}} = 132 \pm 13$ (see Ch. IX, section III-A of Ref.~\onlinecite{abragam1961principles}).}  This has led us to a different approach, described below.

In GaAs, where the conduction-band minimum is at the $\Gamma$ point, we find accurate values of the hyperfine parameters (see Sec.~\ref{sec:numevalcbgaas}).
In contrast, in silicon, where the conduction-band minima are off zone center, we find hyperfine parameters that do not agree with experimental results. 
Therefore, we have evidence that the DFT procedure used here is more accurate for the $\Gamma$-point ($\mathbf{k}=0$) Bloch functions than for Bloch functions at other points in the Brillouin zone.  
Because the point-group symmetry at the $\Gamma$ point is the same as that of the full crystal (as opposed to a subgroup of the crystal point group when $\mathbf{k} \ne 0$), the states at the $\Gamma$ point have higher symmetry than the states at finite $\mathbf{k}$.
Since the basis set used in the Elk code consists of atomic states, which transform according to representations of the full rotation group, it is plausible to expect that the $\Gamma$-point states are more accurate than the states at finite $\mathbf{k}$.
In contrast to the direct DFT calculations at the band extrema described above, here we now use $\textbf{k}\cdot\textbf{p}$ theory to calculate the wave functions at any finite $\mathbf{k}$, starting from the wave functions calculated with DFT at the $\Gamma$ point (``DFT+$\mathbf{k}\cdot\mathbf{p}$'').

To implement DFT+$\mathbf{k}\cdot\mathbf{p}$, we use the experimentally determined values for the $\textbf{k}\cdot\textbf{p}$ matrix elements and energy gaps presented first by Cardona and Pollak \cite{cardona1966energy} and then extended by Richard et al. \cite{richard2004energy} and diagonalize the $\textbf{k}\cdot\textbf{p}$ matrix to determine the correct linear combination of $\mathbf{k}=0$ Bloch amplitudes to describe the states at the conduction-band minima.
We then extract the Kohn-Sham orbitals at the $\Gamma$ point ($\mathbf{k}=0$) and take the appropriate linear combination and (after projecting onto the $s$-component) follow the procedure outlined in Sec.~\ref{sec:numevalcbgaas} for the conduction-band states of GaAs.
Although $\textbf{k}\cdot\textbf{p}$ theory is perturbative, and improves as $k\rightarrow 0$, in Refs.~\onlinecite{cardona1966energy} and \onlinecite{richard2004energy} the entire band structure is shown to be accurately reproduced using these $\textbf{k}\cdot\textbf{p}$ matrix elements and energy gaps.
Therefore, using the matrix elements provided in these references should be sufficient for calculations at the conduction-band minima of silicon. 

The DFT+$\mathbf{k}\cdot\mathbf{p}$ procedure yields $\eta^{\mathrm{Si}} = 88$, which is a factor of $\sim 2$ different from the result ($\eta^{\mathrm{Si}} = 160$) found above for a calculation of the Bloch functions directly at the conduction-band minima.
Furthermore, this DFT+$\mathbf{k}\cdot\mathbf{p}$ result is approximately consistent with the Korringa-relaxation-rate and Knight-shift measurements of Ref.~\onlinecite{sundfors1964nuclear}, $\eta^{\mathrm{Si}} = 100 \pm 10$ (see Appendix \ref{app:Knight} for a discussion of the Knight shift).
This level of consistency suggests that DFT+$\mathbf{k}\cdot\mathbf{p}$ can be useful to perform accurate calculations in materials where the band extrema are not situated at the $\Gamma$ point. The  agreement with experimental observations is also consistent with the assumption of small anisotropic corrections to the Fermi contact interaction. However, since the reduction of $\eta^{\mathrm{Si}}$ is due to a significant $s$-$p$ hybridization, it would still be interesting to assess the role of anisotropy. A proper account of these effects would require applying the methods discussed here to the full bulk states (instead of their $s$ component), but should also take into account the specific nanostructure, e.g., the predominant valley states.\footnote{
We note that in silicon nanostructures (e.g.~quantum dots) the relevant eigenstates are formed by taking linear combinations of the different bulk valley states (and potentially spin states, if spin-orbit coupling is relevant).
In these systems, the anisotropic hyperfine coupling may be significant in comparison to the contact piece.  
If the specific linear combinations making up the eigenstates of a given nanostructure are well understood, the theory presented here can be applied to calculate the appropriate hyperfine tensor.
Instead of projecting the bulk states onto the $s$ component, the full bulk states could be used, appropriate linear combinations taken, and hyperfine tensor elements computed using Eq.~\eqref{eq:hftensor}}

\subsection{Valence bands of GaAs and silicon}\label{sec:numevalvb}

Because the top of the valence band is fourfold degenerate for GaAs and silicon, a general valence-band Kohn-Sham orbital will be a linear combination of all four states.
To calculate the anisotropic hyperfine parameters for these valence-band states, we extract the values of a Kohn-Sham orbital at the top of the valence band on a uniform grid of positions, and use group-theoretic arguments to reconstruct $\phi_{3/2}(\bf{r})$, the Kohn-Sham orbital that transforms like the state with total angular momentum $J=3/2$, orbital angular momentum $l=1$ and $m_J=3/2$ (see Appendix \ref{app:Group Theory}).
We then use the spherical harmonic expansion [Eq.~\eqref{eq:Blochexpansion}] to obtain the radial functions listed in Eqs.~\eqref{eq:radfncs}, \eqref{eq:radfncs2}, and \eqref{eq:radfncs3}.
We find that only the radial functions for quantum number $l$ up to $l=2$ have significant weight (see Appendix \ref{app:conv}). 

The radial functions are inserted into Eq.~\eqref{eq:M} and the appropriate integrals, $M_{\lambda\lambda'}^j$, are computed numerically.
The integrals from Eq.~\eqref{eq:M} are estimated by setting a cutoff for the upper bound of integration at $R_{\mathrm{max}}=\sqrt{3} a /8$, where $a$ is the cubic lattice constant of the material under consideration and $R_{\mathrm{max}}$ is the radius of the largest non-overlapping spheres, ${\cal{S}}_j$, centered at each nuclear site $j$ (see Fig. \ref{fig:fig2}).
Setting the cutoff to $R_{\mathrm{max}}$ is equivalent to neglecting long-range contributions to the hyperfine interaction. %(see explanation in footnote of section \ref{sec:hyperfinenanostructures}). 
We make a further approximation, in the case of the anisotropic hyperfine parameters, and set $f_{\mathrm{T}}(r)\rightarrow 1$ (or equivalently $r_{\mathrm{T}} \rightarrow 0$) when evaluating the matrix elements $M_{\lambda\lambda'}^j$ from Eq.~\eqref{eq:M}.
This is justified because the relativistic radial functions vanish at the origin for all states except $s$ states and $p$ states with total angular momentum $J=1/2$.\cite{pyykko1973hydrogen,vanLeeuwen1994exact}
The valence-band states can be written as a linear combination of $p$ states with $J=3/2$ and $d$ states [see Eq.~\eqref{eq:VBstates}].
Because the relativistic form is important for $r\lesssim r_{\mathrm{T}}$ and the valence-band states vanish at the origin and vary on the scale of a Bohr radius, $a_B$, corrections to the relativistic form are suppressed by $r_{\mathrm{T}}/a_B\lesssim 10^{-3}$. 
Finally, we verify that the computed values of $M_{\lambda\lambda'}^j$ have converged with respect to the parameters listed in Appendix \ref{app:conv} to within 2\% of their asymptotic values (see Appendix \ref{app:conv} for details).

In Refs.~\onlinecite{fallahi2010measurement, chekhovich2013element, vidal2016hyperfine, prechtel2016decoupling}, the ratio of the Overhauser shifts of electrons and holes in GaAs quantum dots is measured. 
From the results of these measurements, the authors conclude $A_{\parallel}/A \sim 10\%$ in GaAs, roughly consistent with the results presented here (see Table \ref{tab:hfparamsVB}). 
In Ref.~\onlinecite{maurand2016cmos}, $T_2^{\ast}$ times have been measured for a hole-spin qubit defined in a silicon complementary metal-oxide-semiconductor (CMOS) quantum dot. 
It is not clear which mechanism limits $T_2^{\ast}$ in these experiments. However, if the coherence times were limited by the hyperfine interaction, the measured $T_2^{\ast}$ times would be consistent with the silicon hyperfine constants presented here (see Appendix~\ref{app:T2s} for details).

\section{Conclusions}\label{sec:conclusions}
We have calculated the hyperfine parameters for the conduction and valence bands of GaAs and silicon using the Kohn-Sham orbitals from an all-electron DFT code (Elk), fully accounting for the relativistic form of the hyperfine coupling, and in the case of silicon, we have introduced and employed an expanded DFT+$\textbf{k}\cdot\textbf{p}$ procedure. 

For the conduction band of GaAs, our results for $\eta^j$ are consistent with the accepted values from Paget et al.~(Ref.~\onlinecite{paget1977low}) and with measurements of the Knight shifts in GaAs quantum wells.\cite{khandelwal1998optically, desrat2013dispersive}
In silicon, our results are roughly consistent with measurements of the Korringa relaxation times and measurements of the Knight shift\cite{sundfors1964nuclear} when we use the DFT+$\textbf{k}\cdot\textbf{p}$ procedure (see Table \ref{tab:eta}).

In the procedure used here, we have accounted for $d$-orbital hybridization in the valence-band states of GaAs.
Similar to the analysis presented in Ref.~\onlinecite{chekhovich2013element}, we find that this $d$-orbital hybridization leads to the Ga nuclear spins (and not the As nuclear spins) in GaAs having a substantial transverse hyperfine coupling ($A^{i}_{\perp} \sim A^{i}_{\parallel}$).
However, while the results of Ref.~\onlinecite{chekhovich2013element} (combined with their interpretation) suggest that heavy holes in a GaAs quantum dot may experience a significant in-plane Overhauser field, we find that the total Overhauser field experienced by a heavy hole in a GaAs quantum dot will point predominantly along the dot growth direction, even for an unpolarized nuclear-spin system.  This anisotropy is a consequence of the stronger hyperfine coupling to the As nuclear spins relative to the Ga nuclear spins: $A^{^{75}\mathrm{As}}_{\parallel} \gg A^{^{69}\mathrm{Ga}/^{71}\mathrm{Ga}}_{\perp}$. 
This finding is consistent with measured heavy-hole spin relaxation times,\cite{gerardot2008optical} transverse Overhauser-field measurements,\cite{prechtel2016decoupling} and measurements of tunneling between spin-resolved Landau levels in a two-dimensional hole gas.\cite{klochan2015landau}
Moreover, we find hyperfine constants that are roughly consistent in magnitude with conclusions drawn in Refs.~\onlinecite{fallahi2010measurement, chekhovich2013element, vidal2016hyperfine, prechtel2016decoupling} from measurements of the ratio of the heavy-hole to electron Overhauser fields.
Additionally, if $A_{\perp}^i$ has a significant magnitude only for the Ga nuclear spins, then in nanostructures (quantum dots or quantum wells) with confined heavy holes and a magnetic field along the growth direction, only the Ga nuclear spins can be dynamically polarized (along the growth direction).
Alternatively, if light holes are confined to similar nanostructures, all nuclear spins can be dynamically spin polarized (see Appendix \ref{app:LHHam}).
Therefore, an additional consequence of the hyperfine constants calculated here is that a larger Overhauser field can be generated if light holes are used to dynamically spin polarize the nuclear spins in GaAs instead of heavy holes. 
For silicon, our results are consistent with $T_2^{\ast}$ measurements made in CMOS hole-spin quantum dots.\cite{maurand2016cmos}
Moreover, in contrast to GaAs, where the hyperfine coupling strength for holes is roughly an order of magnitude smaller than that of electrons ($A_{\parallel}/A \sim 0.1$), in silicon, we find that the hyperfine coupling strengths for holes and electrons are comparible ($A_{\parallel} \sim A$). 

For holes, experiments (including Overhauser-shift and $T_2^{\ast}$ measurements) often only provide indirect measurements of the hyperfine interaction.
For example, extracting the hyperfine parameters from Overhauser-shift measurements requires knowledge of the hole envelope functions, the degree of spin polarization of the nuclear spins, and isotopic alloying disorder.
Measuring the hole hyperfine coupling directly [e.g., through hole-spin echo envelope modulations (HSEEM)\cite{philippopoulos2019hole}] could instead provide a direct and unambiguous measurement of the hyperfine tensor matrix elements, allowing a direct comparison to the theoretical results presented here.  

The method explored here combines DFT, $\mathbf{k}\cdot\mathbf{p}$ theory, and group theory to arrive at an approximate description of the crystal Bloch functions and not only the electronic density.
As demonstrated for the conduction band of silicon, $\mathbf{k}\cdot\mathbf{p}$ theory can be crucial in accurately calculating the Bloch functions away from $\mathbf{k} = 0$. 
The DFT+$\mathbf{k}\cdot\mathbf{p}$ procedure introduced here can therefore be important to understand properties of other materials that have band extrema at finite $\mathbf{k}$.
These materials include graphene, nanotubes, Weyl semimetals, and transition metal dichalcogenides. 
Furthermore, the wave function (including the phase) at all points in the Brillouin zone is required, for example, to evaluate topological invariants (such as Chern numbers). Therefore, DFT+$\mathbf{k}\cdot\mathbf{p}$ might be important in determining topological invariants and cataloguing different topological phases of materials.\cite{bradlyn2017topological, vergniory2019complete}
More generally, this method can be applied to obtain an approximate description of the electronic wave functions for semiconductor systems.
These systems include quantum wells, quantum dots, and defect centers in diamond.
The electronic wave function can be used to calculate relevant quantities in these systems, including, but not limited to hyperfine interactions, spin-orbit interactions, and transition dipole matrix elements.

 \acknowledgments

We are grateful to D.~G.~Austing, H.~Hirayama, M.~Korkusinski, A.~S.~Sachrajda, and S.~A.~Studenikin for helpful discussions.  WAC and PP acknowledge support from Natural Sciences and Engineering Council (NSERC), Canadian Institute For Advanced Research (CIFAR),  Fonds de recherche du Québec – Nature et technologies (FRQNT), and the Gordon Godfrey Bequest. SC acknowledges support from the National Key Research and Development Program of China (Grant No.~2016YFA0301200), NSFC (Grants No.~11574025, No.~11750110428, and No.~1171101295), and NSAF (Grant No.~U1930402).

 %However, since In and Sb have a larger core charge than Ga and As, we expect the renormalization of the orbitals to be less important in InSb than in GaAs. We therefore expect the values of Paget et al.\cite{Paget} to be valid only up to a factor of order 1.%

%Later experiments measured the Knight shift and nuclear relaxation time, $T_1$, for the $^{71}\mathrm{Ga}$ site in GaAs in the quantum Hall regime near a filling factor $\nu=1$.\cite{barrett} A value of the hyperfine contact hyperfine constant is however not possible to extract because of uncertainties in the envelope function. The accuracy of our calculation for the Ga site indicates that our result for the As site should also be accurate. The accuracy of the DFT calculation at  conduction-band minimum of GaAs, situated at the $\Gamma$ point, provides evidence that the DFT calculations can also be trusted at the $\Gamma$ point of silicon.

\appendix

\section{$p$-$d$ hybridization of the valence-band states}\label{app:pd}
In the past, calculations of hole hyperfine constants have been performed by approximating the Bloch amplitudes with atomic $p$ functions.\cite{fischer2008spin, testelin2009hole}
Although they are the simplest states that respect the crystal symmetries, $p$ states are not general enough to completely describe the valence-band Bloch functions (see, for example, the discussion in Sec.~\ref{sec:vbhyperfineparameters} or Refs.~\onlinecite{chadi1976angular, boguslawski1994atomic, diaz2006electronic, chekhovich2013element,machnikowski2019hyperfine}).
The procedure described here allows us to calculate the weight of higher angular momentum states, namely $d$ states, in the valence-band Bloch functions. We can quantify the contribution of the $p$ and $d$ states to the valence-band state around each atom as 
\begin{equation}
w_{\lambda}^j=\frac{\int^{R_{\mathrm{max}}}_{0}\left|R^{j}_{\lambda}(r)\right|^2r^2dr}{\sum_{\lambda', j}\int^{R_{\mathrm{max}}}_{0}\left|R^{j}_{\lambda'}(r)\right|^2r^2dr},
\end{equation} 
for ${\lambda} \in \{p,d,d'\}$. The results are displayed in Table \ref{tab:pdhyb}. The weight of the $p$-orbital increases with the electronegativity of the nucleus (see Table \ref{tab:pdhyb}).
A basic estimate of the electronegativity of the nuclei is given from the effective nuclear charge experienced by the valence electrons of the free atoms calculated from Hartree-Fock theory.\cite{clementi1963atomic}
In GaAs, the As atom, $Z_{\mathrm{eff}}=7.4492$,\cite{clementi1963atomic} has almost pure $p$ symmetry, while the Ga atom, $Z_{\mathrm{eff}}=6.2216$,\cite{clementi1963atomic} has an admixture of $p$ and $d$ symmetries. 
As explained in Sec.~\ref{sec:keyresults}, the higher electronegativity of the As atom suggests a more spherically symmetric potential, and consequently, a weaker $p$-$d$ hybridization (see Fig.~\ref{fig:fig2} and Table~\ref{tab:pdhyb}).
In the case of GaAs, this $p$-$d$ hybridization leads to non-Ising corrections to the heavy-hole hyperfine coupling for the Ga nuclear spins.

Additional evidence for this explanation can be found in Ref.~\onlinecite{chadi1976angular}.
In this reference, the Bloch functions for various zincblende compounds are calculated (using empirical pseudopotentials).
From these calculations, the author concludes that, for the studied materials, as the crystals become more ionic, the $d$-orbitals on the cationic site become more important, and the wave function in the vicinity of the anionic site becomes more $p$-like, consistent with the reasoning provided above.
We also note that, although the $d'$ orbital is allowed by symmetry (see Appendix~\ref{app:Group Theory}), its contribution to the valence-band states vanishes within the accuracy of this procedure. 

Bogus\l{}awski and Gorczyca, Ref.~\onlinecite{boguslawski1994atomic}, have also projected the GaAs valence-band wave functions onto the spherical harmonics. They used the empirical pseudopotential method to obtain the wave functions. These wave functions are then expanded in terms of $p$ and $d$ spherical harmonics centered at each atom $j$. They report results for the $p$ and $d$ contributions to the states from each site. The results of Ref.~\onlinecite{boguslawski1994atomic} are roughly consistent with our own. They also suggest that the contribution of the $d'$ orbitals is relatively small when compared to the $p$ and $d$ orbitals (see Table \ref{tab:pdhyb}). 

Other works using empirical pseudopotentials\cite{chadi1976angular} and tight-binding theory\cite{diaz2006electronic} have also found significant $p$-$d$ hybridization of the Bloch amplitudes near the Ga sites in GaAs.
These works have produced results in rough agreement with the results of Ref.~\onlinecite{boguslawski1994atomic} presented in Table \ref{tab:pdhyb}.

\begin{table}
 \begin{ruledtabular}
\begin{tabular}{lcccc}
 & \multicolumn{2}{c}{present work} & \multicolumn{2}{c}{Ref.~\onlinecite{boguslawski1994atomic}} \\

   atom ($j$) & $w^j_p$ & $w^j_d$  & $w^j_p$ & $w^j_d$   \\
   \hline
   $\mathrm{Ga}$ & $0.13$ & $0.07$ & $0.15$ & $0.20$ \\   
   $\mathrm{As}$ & $0.79$ & $0.01$ & $0.62$ & $0.03$  \\ 
   $\mathrm{Si}$ & $0.43$ & $0.07$  & - & -  \\   
\end{tabular}\caption{Weights, $w_{\lambda}^j$, of the $p$ and $d$ contributions for valence-band states for each atom in GaAs and silicon. In Ref.~\onlinecite{boguslawski1994atomic} silicon is not studied. The weight $w^j_{d'}=0$ within the accurary of the present procedure, which is consistent with the results reported in Ref.~\onlinecite{boguslawski1994atomic}. }\label{tab:pdhyb}
 \end{ruledtabular}
 \end{table}

\section{Comparison with experimental results}\label{app:comparison}

\subsection{Knight shift}\label{app:Knight}
The Knight shift, $K_l$, is the shift in magnetic resonance frequency of an isotope at site $l$ due to the average field $\left<\bm{h}_l\right>$.\cite{abragam1961principles}
Measurements of the Knight shift can be used to characterize the hyperfine interaction for electrons confined to a given nanostructure.  

Measurements of the Knight shift have been made in quantum Hall states of GaAs.\cite{khandelwal1998optically, desrat2013dispersive}
For non-interacting $s$-like electrons (such as those in the conduction-bands of GaAs and silicon) in a quantum well with fully spin-polarized electrons, the Knight shift,
\begin{equation}\label{eq:knightshift}
K_l = \frac{v_0 A^{i_l}}{2 h}\left|F(z_l)\right|^2 n,
\end{equation}
is proportional to the hyperfine constant $A^{i_l}$. In Eq.~\eqref{eq:knightshift}, $h$ is Planck's constant, $F(z)$ is the quantum-well envelope function, and $n$ is the sheet density of electrons in the quantum well. In Ref.~\onlinecite{khandelwal1998optically}, the Knight shift for nuclei at the center of a GaAs quantum well was measured using optically pumped nuclear magnetic resonance in three different samples in the $\nu = 1/3$ fractional quantum Hall state (having a fully spin-polarized ground state).
For a symmetric quantum well with infinite barriers, the largest Knight shift occurs directly in the center of the well, and is proportional to $\left|F(z=L/2)\right|^2 = 2/L$, where $L$ is the well width. 
Using this value for the envelope function, the hyperfine coupling was extracted from the Knight-shift measurement and a value of $A_c^{^{71}\mathrm{Ga}}=v_0 A^{^{71}\mathrm{Ga}}/h = (4.5 \pm 0.2) \times 10^{-13}\,\mathrm{cm}^{3}/\mathrm{s}$ was reported.
This value can be converted into a value for $\eta^{j_l}$ (for atom $j_l$ at site $l$) using Eq.~\eqref{eq:contacthfparameter}, and is presented in Table \ref{tab:eta}. More recently, Knight-shift measurements have been made in GaAs in the quantum Hall regime, close to a filling factor $\nu = 1$.\cite{desrat2013dispersive} The results for the Knight shifts for $^{69} \mathrm{Ga}$ and $^{75}\mathrm{As}$ relative to that of $^{71} \mathrm{Ga}$ (plotted in Fig. 1 of Ref.~\onlinecite{desrat2013dispersive}) can be combined with the Knight-shift measurement of Ref.~\onlinecite{khandelwal1998optically} and Eq.~\eqref{eq:knightshift} to obtain values for the hyperfine constants for $^{69} \mathrm{Ga}$ and $^{75}\mathrm{As}$. The values of $\eta^{j_l}$ obtained from these measurements are consistent with our calculated values (see Table \ref{tab:eta}).

The Knight shift has also been measured in $n$-doped bulk silicon samples.\cite{sundfors1964nuclear}
The extracted hyperfine parameter is $\eta^{\mathrm{Si}}=100\pm10$, approximately consitent with our calculated value of $\eta^{\mathrm{Si}}=88\pm1$ (see Table \ref{tab:eta}).

\subsection{Hole-spin coherence times}\label{app:T2s}

The hyperfine field can limit coherence times for electrons \cite{} or holes \cite{maurand2016cmos} trapped in nanostructures. Recently, Maurand et al.,\cite{maurand2016cmos} measured the coherence time, $T_2^{\ast}=\left(59 \pm1\right) \ \mathrm{ns}$, of a hole-spin qubit confined to a CMOS silicon quantum dot.
Under the assumption that there are enough nuclear spins interacting with the hole spin that the hyperfine-field value will be Gaussian distributed, we can estimate the coherence time for the heavy-hole spin using \cite{fischer2008spin}
\begin{equation}\label{eq:T2star}
\frac{1}{2(T_2^{\ast})^2}\approx\frac{1}{4N}\sum_i g_{i}I_i(I_i+1)(A^{i}_{\parallel})^2,
\end{equation} 
where $g_i$ is the abundance of isotope $i$ having nuclear spin $I_i$, and $N$ is the number of nuclear spins in the nanostructure.
From the quantum dot level-spacing from Maurand et al.,\cite{maurand2016cmos} we estimate $N\sim10^3$, assuming a spherical quantum dot.
Calculating $T_2^{\ast}$ from Eq.~\eqref{eq:T2star} using our result for $A^{\mathrm{Si}}_{\parallel}$, $I_i=1/2$, and the natural abundance of $^{29}\mathrm{Si}$ ($g_{^{29}\mathrm{Si}} = 4.7 \%$), we find $T_2^{\ast}$ to be on the order of $100 \, \mathrm{ns}$.
Our estimate of $T_2^{\ast}$ is therefore of the same order as the measured value.

\section{Group theory and projection operators}\label{app:Group Theory}

To reconstruct the heavy-hole and light-hole states from an arbitrary linear combination of these four states, we use the projection operator technique \cite{dresselhaus2008group} from group theory.

The states at the top of the valence band of group IV and III-V semiconductors transform according to the $\Gamma_8$ representation of the tetrahedral double group, $T_d$ (or equivalently the $\Gamma_8^+$ representation of the $O_h$ double group). \cite{dresselhaus2008group} We start by constructing the $\Gamma_8$ representation and then show how it can be used along with the projection operators to determine states that will contribute to the partial-wave expansion of the heavy-hole and light-hole states.

\subsection{Construction of the $\Gamma_8$ representation}

A known basis for the $\Gamma_8$ representation of the tetrahedral double group, $T_d$, is the set of four $J=3/2$ angular momentum eigenstates with $l=1$ (see Table D.1. p.~522 in Ref.~\onlinecite{dresselhaus2008group}).
In the $\ket{J,l,m_J}$ basis, where $J$ represents the total angular momentum, $l$ gives the orbital angular momentum, and $m_J$ is the angular momentum projected onto the relevant quantization axis (the $z$-axis, e.g.~$[001]$), these states are $\ket{3/2, 1, \pm 3/2}$, which transform like the heavy-hole states, and $\ket{3/2, 1, \pm 1/2}$, which transform like the light-hole states. According to the definition of basis vectors, \cite{dresselhaus2008group} 
\begin{equation}\label{eq:basisdef}
 {\cal{O}}_i \ket{3/2,1,m_J}=\sum^{3/2}_{m_J'=-3/2}[D^{(\Gamma_8)}({\cal{O}}_i )]_{m_Jm_J'}\ket{3/2,1,m_J'},
\end{equation}
where ${\cal{O}}_i$ $\in$ $T_d$ is a symmetry operation and $D^{(\Gamma_8)}({\cal{O}}_i )$ is the $\Gamma_8$ representation of the ${\cal{O}}_i$ symmetry. 

Using the orthonormality of the basis states, we can construct the $\Gamma_8$ representation matrices as 
\begin{equation}\label{eq:repelements}
[D^{(\Gamma_8)}({\cal{O}}_i )]_{m_Jm_J'}=\bra{3/2,1,m_J'}{\cal{O}}_i\ket{3/2,1,m_J},
\end{equation}
for all symmetry operations ${\cal{O}}_i$ $\in$ $T_d$. Furthermore, we have
\begin{equation}\label{eq:Wignerdef}
\bra{3/2,1,m_J'}{\cal{O}}_i\ket{3/2,1,m_J}=\sigma({\cal{O}}_i)W^{3/2}_{m_Jm_J'}(a_i, b_i, c_i),
\end{equation}
where $\sigma({\cal{O}}_i)=1$ if the operation is a pure rotation, $\sigma({\cal{O}}_i)=(-1)^l$ if the operation involves an inversion, and $W^J(a,b,c)$ is the $J^\mathrm{th}$ Wigner D matrix. The angles  $a_i$, $b_i$ and $c_i$ are the symmetry-dependent Euler angles,  where $a_i$ is an initial rotation around the $z$-axis, $b_i$ a subsequent rotation around a perpendicular axis, labeled $y$ ($[010]$) and $c_i$ is the final rotation around the $z$-axis (these angles can be found for the different symmetry operations ${\cal{O}}_i$ $\in$ $T_d$ in Table I of Ref.~\onlinecite{fox1970construction}). Inserting Eq.~\eqref{eq:Wignerdef} into Eq.~\eqref{eq:repelements}, we can construct the $\Gamma_8$ representation of the $T_d$ double group, $\{D^{(\Gamma_8)}({\cal{O}}_i)\}$.

\subsection{Projection operators}\label{app:GTPO}
Since each valence-band state transforms like one of the four $\ket{3/2,1,m_J}$ states, we label each state by $m_J$. This labeling is consistent with the notation developed above.
We now define the projection operators and show how to use them to construct the heavy-hole and light-hole states. For the $\Gamma_8$ basis states, the projection operators $\hat{P}_{m_Jm_J'}$, are defined by the equation 
\begin{equation}
\hat{P}_{m_Jm_J'}\ket{m_J'}=\ket{m_J},
\end{equation}
where $m_J$ and $m_J'$  run over the four basis states of the $\Gamma_8$ representation.
Under the $\Gamma_8$ representation, the projection operators are written as \cite{dresselhaus2008group}
\begin{equation}\label{eq:projop}
\hat{P}_{m_Jm_J'}=\frac{1}{12}\sum_{i}\{[D^{(\Gamma_8)}({\cal{O}}_i)]^{-1}\}_{m_Jm_J'}^{\ast}{\cal{O}}_i,
\end{equation}
where the numerical prefactor comes from the ratio of the order of the $\Gamma_8$ representation to the order of the double group $T_d$.

Because the projection operators are linear, $\hat{P}_{3/2,3/2}$ can retrieve the component of any state that transforms like $\ket{3/2}$ under the symmetry operations of the double group $T_d$. Therefore, by systematically applying the $\hat{P}_{3/2,3/2}$ projection operator to the 6 $p$-states ($l=1$) and the 10 $d$-states ($l=2$), we can calculate the $d$-orbital hybridized heavy-hole state $\ket{3/2}$. The result is

\begin{align}\label{eq:HHup}
\braket{r}{3/2}=&R_p(r)\ket{1,1}\ket{+}-iR_d(r)\ket{2,-1}\ket{+}\\
&-iR_{d'}(r)\ket{2,0}\ket{-}, \nonumber
\end{align}
where $r$ is a radial coordinate, $R_{\lambda}(r)$ are real radial functions and we have used the basis of states $\ket{l,m}\ket{\sigma}$. The basis vectors $\ket{l,m}\ket{\sigma}$ are related to $\ket{J,l,m_J}$ basis vectors by the Clebsch-Gordon coefficients. 
We then construct the other three states by applying the projection operators $\hat{P}_{m_J,3/2}$ to the state from Eq.~\eqref{eq:HHup} 

\begin{equation}\label{eq:VBstates}
\begin{split}
&\begin{split}
\braket{r}{-3/2}=&R_p(r)\ket{1,-1}\ket{-}+iR_d(r)\ket{2,1}\ket{-}\\
&+iR_{d'}(r)\ket{2,0}\ket{+},
\end{split}\\
&\begin{split}
\braket{r}{+1/2}=&R_p(r)\left(\sqrt{\frac{2}{3}}\ket{1,0}\ket{+}+\sqrt{\frac{1}{3}}\ket{1,1}\ket{-}\right)\\
&-i\tilde{R}_1(r)\ket{2,2}\ket{+}-i\tilde{R}_{2}(r)\ket{2,-2}\ket{+}\\
&-i\frac{R_{d}(r)}{\sqrt{3}}\ket{2,-1}\ket{-},
\end{split}\\
&\begin{split}
\braket{r}{-1/2}=&R_p(r)\left(\sqrt{\frac{2}{3}}\ket{1,0}\ket{-}+\sqrt{\frac{1}{3}}\ket{1,-1}\ket{+}\right)\\
&+i\tilde{R}_1(r)\ket{2,-2}\ket{-}+i\tilde{R}_{2}(r)\ket{2,2}\ket{-}\\
&+i\frac{R_{d}(r)}{\sqrt{3}}\ket{2,1}\ket{+},
\end{split}\\
\end{split}
\end{equation}
where $\tilde{R}_1(r)=\sqrt{1/3}R_d(r)+\sqrt{1/2}R_{d'}(r)$ and $\tilde{R}_2(r)=\sqrt{1/2}R_{d'}(r)-\sqrt{1/3}R_{d}(r)$.

Finally, we note that we also enforce 
\begin{equation}
\Theta \braket{r}{3/2} = e^{i\phi_0} \braket{r}{-3/2}
\end{equation}
where $\Theta$ is the time-reversal operator and $\phi_0$ is a global phase. This equation restricts the relative phases of the $p$ and $d$ parts of the wave functions to be as shown in Eqs.~\eqref{eq:HHup} and \eqref{eq:VBstates}. We also note that we have omitted the $\textbf{k}_{\nu}$ quantum number for the valence-band states since the valence-band maximum for group IV and III-V semiconductors is situated at the $\Gamma$ point, where $\textbf{k}=0$.

The advantage of applying the projection operators to atomic orbitals is that the symmetry of the states can be easily identified.
For example, in the case of the valence-band states of GaAs and silicon, the deviation from pure $p$ symmetry (and the $d$-orbital hybridization) can be easily understood by using the projection operators (as described above) to write the states as in Eqs.~\eqref{eq:HHup} and \eqref{eq:VBstates}.
We note that the group theory projection operators can also be applied directly to wave functions defined numerically on a grid of points by applying the symmetry operators ${\cal{O}}_i$ [see Eq.~\eqref{eq:projop}] directly to the coordinates (for an implementation, see Ref.~\onlinecite{philippopoulosgithub}). 

\section{Light-hole hyperfine Hamiltonian}\label{app:LHHam}

Projecting the hyperfine matrix [Eq.~\eqref{eq:hfHamnoEnv}] onto the light-hole subspace results in
\begin{align}
\mathrm{H}^{l,A}_{\mathrm{LH}} &=\frac{1}{2}\bigg[\left(\frac{1}{3}A^{i_l}_{\parallel} - 4 A^{i_l}_{\perp}\right)\sigma_zI^{l,A}_z \label{eq:hfLHA}\\ 
&+\left(\frac{2}{3}A^{i_l}_{\parallel} + A_{\perp}^{i_l}\right)\left(\sigma_xI^{l,A}_x+\sigma_yI^{l,A}_y\right)\bigg] \label{eq:hfLHA2}, 
\end{align}
where the Pauli matrices, $\sigma_{\alpha}$, act in the light-hole subspace. 
This hyperfine matrix is given for an isotope $i_l$ located at an $A$ site labeled by $l$ (see Sec.~\ref{sec:coordinates}).
In contrast to the heavy-hole hyperfine matrix, the light-hole hyperfine matrix is invariant under  $A^{xx} \rightarrow A^{yy}$ and $A^{yy} \rightarrow A^{xx}$. 
Therefore the hyperfine matrix for $A$ sites is equivalent to the hyperfine matrix for $B$ sites. 
In addition, the logitudinal and transverse light-hole hyperfine couplings depend on both $A_{\parallel}$ and $A_{\perp}$ [see Eqs.~\eqref{eq:hfLHA} and \eqref{eq:hfLHA2}]. 
Therefore, in contrast to heavy holes, even when $A_{\perp} \ll A_{\parallel}$ (e.g. for the As site in GaAs), the transverse light-hole hyperfine coupling is of the same order as the longitudinal hyperfine coupling [see Eqs.~\eqref{eq:hfLHA} and \eqref{eq:hfLHA2}].  

\section{$s$-like Kohn-Sham orbitals}\label{app:snorm}
The $s$-like states that contribute to the conduction-band minimum of GaAs are `almost purely $s$-like,' which we take to mean
\begin{equation}
\frac{\sum_{l>0} N^{\nu}_l}{N^{\nu}_0}<10^{-3},
\end{equation}
where
\begin{equation}\label{eq:defNl}
N^{\nu}_l = \sum_{m=-l}^l\sum_{\sigma,j}\int_0^{R_\mathrm{max}}\left|R^{j\nu}_{lm\sigma}\right|^2r^2dr.
\end{equation}
In Eq.~\eqref{eq:defNl} $R_\mathrm{max} = \sqrt{3} a /8$, where $a$ is the lattice constant. $R_\mathrm{max}$ is half the distance between nearest-neighbor atoms in the crystal. 
 
For $s$-like orbitals the contact hyperfine interaction dominates. Since the contact hyperfine constants are  determined by the integral $\left<\left|R^j_s(r)\right|^2\right>_{\delta_{\mathrm{T}}}$, we present here the procedure used to evaluate this integral. Since the relativistic $s$-like radial function has a power-law divergence at the origin,\cite{pyykko1973hydrogen,vanLeeuwen1994exact} we fit the points that are within a distance of $10r_{\mathrm{T}}^j$ from each atom with a power law and evaluate the integral 
\begin{equation}
\left<\left|R^j_s(r)\right|^2\right>_{\delta_{\mathrm{T}}}^{\mathrm{in}}=\int_0^{10r_{\mathrm{T}}^j} \left|R_{\mathrm{fit},s}^{j}(r)\right|^2 \delta_{\mathrm{T}}(r)r^2dr,
\end{equation}
where $R_{\mathrm{fit},s}^{j}(r) = \Lambda r^{-\xi}$ is the best fit function to the radial part of the Kohn-Sham orbital, with $\Lambda$ and $\xi$ being fit parameters. We then use a Riemann sum to evaluate the integral for all points $10r_{\mathrm{T}}^j<r_n<100r_{\mathrm{T}}^j$,
\begin{equation}
\left<\left|R^j_s(r)\right|^2\right>_{\delta_{\mathrm{T}}}^{\mathrm{out}}=\sum_{r_n=10r_{\mathrm{T}}^j}^{100r_{\mathrm{T}}^j}\left|R_n^j\right|^2\delta_{\mathrm{T}}(r_n)r_n^2 \Delta,
\end{equation}
where $r_n$ is the set of points where the radial function $R_s^j(r)$ is sampled, $R_n^j=R_s^j(r_n)$, and $\Delta = r_{n+1}-r_n$. We then approximate 
\begin{equation}\label{eq:appinout}
\left<\left|R^j_s(r)\right|^2\right>_{\delta_{\mathrm{T}}} \approx \left<\left|R^j_s(r)\right|^2\right>_{\delta_{\mathrm{T}}}^{\mathrm{in}}+\left<\left|R^j_s(r)\right|^2\right>_{\delta_{\mathrm{T}}}^{\mathrm{out}}.
\end{equation}
In Eq.~\eqref{eq:appinout} we have taken contributions to the integral [Eq.~\eqref{eq:defav}] to be negligible for $r > 100 r_{\mathrm{T}}$.
This approximation is justified because the scale at which the weighting function in the integral [$\delta_{\mathrm{T}}(r)$] decays is given by $r_{\mathrm{T}}$ [see Eq.~\eqref{eq:deltaT}].

\section{Convergence criteria}\label{app:conv}

 For each parameter $p$ (e.g.~$p$ can be the density of $k$ states at which the DFT calculation is performed), we construct $\alpha^{j}(p)$, $\alpha \in \{\eta, h_{\perp}, h_{\parallel}\}$. In other words, we evaluate $\alpha^{j}$ for a range of values of the parameter $p$.
Once $\alpha^{j}$ has been evaluated for multiple values of $p$, we fit $\alpha^{j}(p)$ to a power law of the form
\begin{equation}
\alpha^{j}(p)=\Lambda p^{-\xi}+\alpha^j_0,
\end{equation}
where $\Lambda$, $\xi$ and $\alpha^j_0$ are fit parameters and, in particular, $\alpha^j_0$ is the asymptotic value of $\alpha^j$ as a function of $p$. In all cases, we find that  
\begin{equation}
\frac{\left|\alpha^j(p_{\mathrm{vhq}})-\alpha^j_0\right|}{\alpha^j_0}<e,
\end{equation}
where $p_{\mathrm{vhq}}$ is the `very high quality' value of the paramter $p$, determined by Elk,\cite{elk} and $e=0.01$ for $\alpha=\eta$ and $e=0.02$ for $\alpha \in \{h_{\perp}, h_{\parallel}\}$. This procedure was carried out for the parameters gmaxvr, lmaxvr, nempty, rgkmax, chgexs, swidth,\cite{elk} as well as the number of $k$ points in the first Brillouin zone at which the Kohn-Sham orbitals were found, and the number of points in the unit cell at which the wave functions were extracted.

In addition, we have verified the smallness of the error made in expanding the valence-band states only up to $l=2$ in the spherical harmonic expansion [see Eq.~\eqref{eq:Blochexpansion}].
Specifically, if we define 
\begin{equation}
\mathcal{M}_j(l_{\mathrm{max}})=\sum_{\sigma}\int_{\mathcal{S}_{j}}drd\Omega\frac{\left|\sum^{l=l_{\mathrm{max}}}_{l=0}\sum_{m=0}^{l}R_{lm\sigma}^{j\nu}(r)Y_{lm}(\theta,\phi)\right|^2}{r},
\end{equation}
where the integral is over the sphere ${\cal{S}}_j$ surrounding atom $j$ (see Fig.~\ref{fig:fig2}), we have verified that 
\begin{equation}\label{eq:errormeasureshe}
\frac{\mathcal{M}_j(3)-\mathcal{M}_j(2)}{\mathcal{M}_j(3)}<0.001
\end{equation}
for all atoms $j$ in GaAs and silicon.
Since the hyperfine parameters are calculated from matrix elements of $\mathbf{h}(\mathbf{r}) \sim 1/r^3$ [see Eq.~\eqref{eq:hrSingleParticleMatrix}] and $\mathcal{M}_j(l_{\mathrm{max}})$ is a (diagonal) matrix element of $1/r^3$, Eq.~\eqref{eq:errormeasureshe} should be a good measure of the error made in neglecting terms with $l>2$ when calculating hyperfine constants.

%\begin{acknowledgments}
%\end{acknowledgments}

\bibliography{DFThyperfine}

\end{document}